\documentclass[journal,10pt]{IEEEtran}

\usepackage[utf8]{inputenc}
\usepackage[english]{babel}
\usepackage{amsmath,amssymb,bm,afterpage,capt-of}
\usepackage{graphicx,caption,hyperref,subfig,afterpage,stfloats}
\usepackage{verbatim,fancyvrb}
\usepackage{placeins} 
\usepackage{algorithm,algorithmic}
\usepackage{framed}
\usepackage{booktabs,colortbl,multirow,makecell}
\usepackage{setspace}
\usepackage{tikz,tikz-3dplot}
\usepackage{import}
\usepackage{url}
\usepackage{lipsum}
\newcolumntype{P}[1]{>{\centering\arraybackslash}p{#1}}
\usepackage{amsthm}
\usepackage{amssymb}
\theoremstyle{definition}
\newtheorem{definition}{Definition}[section]
\newtheorem{property}{Property}[section]
\newtheorem{lemma}{Lemma}[section]
\newtheorem{remark}{Remark}[section]

\DeclareMathOperator*{\argmin}{arg\,min}

\DeclareMathOperator*{\argminA}{arg\,min}

\begin{document}

\title{Variable-size Symmetry-based Graph Fourier Transforms for image compression}
\author{Alessandro~Gnutti,~\IEEEmembership{Member,~IEEE}, Fabrizio~Guerrini,~\IEEEmembership{Member,~IEEE}, Riccardo~Leonardi,~\IEEEmembership{Fellow,~IEEE}, and~Antonio~Ortega,~\IEEEmembership{Fellow,~IEEE}%
\thanks{Alessandro Gnutti and Fabrizio Guerrini are with the Department of Information Engineering, CNIT -- University of Brescia, 25123 Brescia, Italy.

Riccardo Leonardi is with the Department of Information Engineering, CNIT -- University of Brescia, 25123 Brescia, Italy, and also with the Department
of Electronics Engineering, University of Rome Tor Vergata, 00133 Roma,
Italy.

Antonio Ortega is with the Department of Electrical and Computer Engineering, University of Southern California, Los Angeles, USA.

Corresponding author: Alessandro Gnutti, e-mail: alessandro.gnutti@unibs.it.
}
}

\markboth{Submitted to IEEE Transactions on Circuits and Systems for Video Technology}%
{Submitted to IEEE Transactions on Circuits and Systems for Video Technology}

\maketitle


\begin{abstract}
Modern compression systems use linear transformations in their encoding and decoding processes, with transforms providing compact signal representations. 
While multiple data-dependent transforms for image/video coding can adapt to diverse statistical characteristics, assembling large datasets to learn each transform is challenging.
Also, the resulting transforms typically lack fast implementation, leading to significant computational costs.
Thus, despite many papers proposing new transform families, the most recent compression standards predominantly use traditional separable sinusoidal transforms. 
This paper proposes integrating a new family of Symmetry-based Graph Fourier Transforms (SBGFTs) of variable sizes into a coding framework, focusing on the extension from our previously introduced 8x8 SBGFTs to the general case of NxN grids.
SBGFTs are non-separable transforms that achieve sparse signal representation while maintaining low computational complexity thanks to their symmetry properties.
Their design is based on our proposed algorithm, which generates symmetric graphs on the grid by adding specific symmetrical connections between nodes and does not require any data-dependent adaptation.
Furthermore, for video intra-frame coding, we exploit the correlations between optimal graphs and prediction modes to reduce the cardinality of the transform sets, thus proposing a low-complexity framework.
Experiments show that SBGFTs outperform the primary transforms integrated in the explicit Multiple Transform Selection (MTS) used in the latest VVC intra-coding, providing a bit rate saving percentage of $\mathbf{6.23\%}$, with only a marginal increase in average complexity.
\textit{A} MATLAB \textit{implementation of the proposed algorithm is available online at} \cite{github}.
\end{abstract}
\begin{keywords}
Image/video compression, multiple transforms, Graph Signal Processing, Graph Fourier Transform, symmetric graphs. 
\end{keywords}


\section{Introduction}
\noindent
Image and video compression aims to encode data efficiently while preserving the original visual quality. Traditionally, linear block transform coding plays a vital role in compression techniques. The rationale underlying this paradigm is to design the transforms in such a way as to provide statistically independent coefficients in the transform domain, which leads to a more effective encoding.

Among the various transforms proposed in the past decades, the 2D type-$2$ Discrete Cosine Transform (DCT-II, or simply DCT) is a widely adopted choice due to its ability to approximate the optimal Karhunen-Loève Transform (KLT) under first-order Markov assumptions \cite{Rao1975}, which are well-suited for natural imaging content. For this reason, the DCT is employed in many popular block-based compression techniques. The DCT has been used in video compression since the ITU/H.261 standard \cite{Bhaskaran1995} up to JVT/H.265 (HEVC) \cite{intraHEVC2012} and the most recent JVT/H.266 (VVC) \cite{vvc2021}. In image compression, the same applies, from the well-established ISO/JPEG standard \cite{jpeg1992} to the recent BPG \cite{bpg}, derived from HEVC intra-frame coding.

However, it is well-known that a fixed transform may not be a suitable model for visual data \cite{Dony-TSP-1995}. Thus, letting the encoder and the decoder share \textit{multiple transforms} in the compression framework helps to adapt to the diverse statistical characteristics of both natural images and residuals.

Considering transforms specific for video coding, one possible approach is to develop effective transforms based on the type of residual data. For instance, Mode-Dependent Transform (MDT) schemes, which involve designing a different KLT for each intra-prediction mode, have been proposed for H.264/AVC \cite{ye2008improved} and for HEVC \cite{takamura2013intra, arrufat2014non}. While MDTs are simple and efficient, they have the limitation of employing a fixed transform for each class of signals, namely, residual blocks deriving from the same mode prediction. However, significant statistical variations within these signal classes persist, so MDTs do not entirely overcome the limitations of using a single transform.

Alternatively, we can adopt a Rate-Distortion Optimized Transform (RDOT) scheme \cite{zhao2011video}. In this scenario, the encoder selects the most appropriate transform for each block by minimizing a defined Rate-Distortion (RD) cost. Finally, the corresponding transform index must be sent to the decoder along with the transform coefficients for reconstruction.

For an RDOT approach to work, one should ideally design a family of transforms that effectively capture the varied visual data statistical characteristics. For example, a representative image (or residual) block training set can be collected, and successively, the blocks can be clustered according to suitable criteria so that a specific KLT can be built for each cluster. 
However, the KLTs are derived from sample covariance matrices, which may not be good estimators for the true covariance of the underlying models, especially when the number of data samples is small \cite{johnstone2009consistency}. Thus, collecting a sufficiently representative dataset to capture statistical variations of signals to design multiple data-dependent transforms is challenging. Moreover, the resulting transforms are typically non-separable and lack fast implementation, leading to significant computational costs.

In light of the aforementioned challenges, data-driven transforms are not commonly used. The latest compression standards \cite{intraHEVC2012, vvc2021} use a straightforward RDOT scheme with a limited number of fixed, computationally efficient, and separable transforms. In HEVC, the Discrete Sine Transform (DST) supplements the DCT, while its successor VVC includes four additional primary transforms besides the DCT-II: the type-$8$ DCT (DCT-VIII), the type-$7$ DST (DST-VII), and their horizontal and vertical combinations. 

Recent research has suggested using Graph Fourier Transforms (GFTs) for image and video coding. This approach is based on Graph Signal Processing (GSP) \cite{Shuman-SPM-2012, Ortega-proceedings-2018}, where graphs are adopted as models in which the nodes identify the signal samples and the edge weights represent the similarity between samples. 
In \cite{fracastoro2019graph}, a block-adaptive scheme is presented for image compression with GFTs, where a graph is constructed for each block signal by minimizing a regularized Laplacian quadratic term used as the proxy for the actual RD cost. 
Also, the authors of \cite{shen2010} introduce Edge Adaptive Transforms (EAT) based on graphs specifically for depth-map compression, and in \cite{wei2015}, the EATs are extended for piecewise-smooth image compression. However, these works are inefficient from an RD perspective because it is necessary to send the graph description for each block. 
The two types of Graph-based transforms (GBTs) for video compression proposed in \cite{egilmez2020} also have limitations: the data-driven approaches suffer from the same disadvantages as the aforementioned KLTs-based solutions, while transforms that explicitly identify image edges require adjusting the transform for each block, leading to increased encoding times.

In the same GSP vein, our previous work \cite{gnutti18} introduced preliminary symmetric graphs specifically for $4{\times}4$ grids. Although the corresponding GFTs demonstrated good approximation abilities, these transforms lacked a fast implementation and were limited to $4{\times}4$ blocks. Then, our following works \cite{Gnutti_icip19, gnutti21} put forward a set of computationally efficient, non-separable 2D GFTs with directional bases named Symmetry-based Graph Fourier Transforms (SBGFTs), which achieve a sparse signal representation and higher image compression performance. In particular, in \cite{Gnutti_icip19}, we presented preliminary results indicating the proposed $8{\times}8$ SBGFTs provide better energy approximation than the $8{\times}8$ DCT. 
In \cite{gnutti21}, we expanded this comparison, showing that the $8{\times}8$ SBGFTs still exhibited energy approximation capabilities superior to those of alternative sets of transforms, such as KLTs and Sparse Orthonormal Transforms (SOTs) \cite{sezer2015sparse}.

To achieve highly efficient computation of the SBGFTs, we constrained the graph structures to be symmetric, following the ideas in \cite{KSLU_ICASSP17, KSLU_TSP19}, in which the authors have shown that fast transform implementations can be achieved by exploiting graph grid symmetries. Our proposed SBGFTs can reduce the complexity to half that of other non-separable approaches. Their complexity even becomes comparable to that of separable transforms when there is more than one symmetry direction in the graph.
Additionally, in our recent work \cite{gnutti2024symmetric}, we proposed an effective approach to reduce the average number of overhead bits required to signal the transform index in a multiple transforms scenario and demonstrated the efficiency of this method experimentally by applying it to the SBGFTs. Naturally, the SBGFTs in \cite{Gnutti_icip19, gnutti21, gnutti2024symmetric} still have the limit of considering only $8{\times}8$ blocks. Since current image and video compression standards use variable-size image blocks, this reduces the potential performance improvement to only a subset of residual blocks.

In this paper, we significantly advance practical SBGFTs for image and video compression by extending the original $8{\times}8$ SBGFTs to the general case of $N{\times}N$ grids. This is achieved by generalizing the algorithm that builds symmetric graphs, describing an efficient strategy for adding symmetrical connections between edges. In particular, our algorithm can consider more potential graph connections compared to the $8{\times}8$ case, enabling the basis functions of the resulting SBGFT to better adapt to the data's statistics.

This generalization to $N{\times}N$ transforms allows us to simulate the performance of a VVC-like coding framework using variable-size SBGFTs.
In particular, we compare the RD performance of the proposed variable-size SBGFTs in the residual domain with the Explicit Multiple Transform Selection (MTS) method of VVC, which 
selects among multiple sinusoidal transforms using the RDOT paradigm. Our study includes different configurations highlighting the benefits of extending to variable-size SBGFTs.
Our results reveal that employing the variable-size SBGFTs framework on residual blocks yields a Bj\o ntegaard's Delta-Rate (BD rate) reduction of $9.29\%$ compared to the variable-size MTS method of VVC, which serves as the baseline. Significantly, when only $8{\times}8$ SBGFTs are used, the BD rate reduction drops to $5.14\%$.

Moreover, we investigate how frequently each graph is chosen for residual blocks obtained from a given prediction mode. Based on statistics collected in our experiments, we propose reducing encoding complexity by selecting optimal subsets of transforms tailored to the prediction modes integrated in the intra-coding of VVC. This solution allows us to achieve a BD rate saving of $6.23\%$ while maintaining the same complexity as the MTS method in VVC. 

The main contributions of this paper are:
\begin{itemize}
    \item We extend the SBGFTs to square blocks of arbitrary size, allowing our algorithm to include more potential graph connections compared to the method restricted to $8{\times}8$ grids.
    \item We test the SBGFTs in a VVC-like framework and show their superior performance with respect to VVC primary transforms in the Explicit MTS configuration.
    \item We propose a strategy for reducing the encoding complexity inherent in the RDOT mechanism when the set of transforms considered is large, by exploiting correlations between optimal graphs and the prediction mode that generated the residual block.
\end{itemize}

The rest of the paper is organized as follows. In Sec.~\ref{sec:preliminaries}, we briefly review GSP concepts and the efficient GFT implementation properties of symmetric graph grids. Sec.~\ref{sec:sbgft} formally defines the $N{\times}N$ SBGFTs, while extensive experimental results are presented in Sec.~\ref{sec:exp}. 
Sec.~\ref{sec:concl} concludes the paper.

\section{Preliminaries}
\label{sec:preliminaries}

\subsection{Background}
\label{subsec:background}

In GSP \cite{Shuman-SPM-2012, Ortega-proceedings-2018}, graph signals are defined on a connected and weighted graph that is usually denoted as $\mathcal{G}=\{\mathcal{V},\mathcal{E},\mathbf{W}\}$.
The graph is built upon a set of vertices (or nodes) $\mathcal{V}={1,2,\ldots, N_V}$, on which the signal samples are located, a set of edges $\mathcal{E}$ connecting the nodes of the graph that carry the signal values, and a weighted adjacency matrix $\mathbf{W}$ containing the edge weights as entries. Therefore, $\mathcal{G}$ establishes a relationship among the signal samples. Graphs can be either directed or undirected. However, we focus solely on undirected graphs without self-loops, i.e., $\mathbf{W}$ is symmetric and has zero-valued diagonal entries. 

Denoting $\mathbf{D}$ the diagonal degree matrix of $\mathcal{G}$, we can define the Laplacian matrix as $\mathbf{L}=\mathbf{D}-\mathbf{W}$. To assess the smoothness of a graph signal $\mathbf{f}$, we use the Laplacian quadratic form: 
\begin{equation}
\label{eq:quadratic_form}
    \mathbf{f}^\top \mathbf{L}\, \mathbf{f} = \sum\limits_{(i,j)\in \mathcal{E}} w_{i,j}\cdot ( f_i - f_j )^2\ ,
\end{equation}
where $w_{i,j}$ are the edge weights, and $f_i$ and $f_j$ are the signal values at nodes $i$ and $j$. 
$\mathbf{f}^\top \mathbf{L}\, \mathbf{f}$ is zero if and only if the signal $\mathbf{f}$ is constant across all vertices, and it is small when $\mathbf{f}$ takes similar values at neighboring vertices.


Spectral representations of graph signals utilize the eigenvalue decomposition of either the adjacency or the Laplacian matrix. Since the graphs are undirected, $\mathbf{L}$ is symmetric, and thus its eigendecomposition gives us  $\mathbf{L}=\mathbf{U}\mathbf{\Lambda}\mathbf{U}^\top$, where $\mathbf{U}$ represents a real-valued orthonormal matrix formed by unit norm eigenvectors of $\mathbf{L}$, and $\mathbf{\Lambda}$ is a diagonal matrix with the corresponding real eigenvalues $\lambda_j$. Thus, the column eigenvectors $\mathbf{u}_j$ of 
$\mathbf{U=[\mathbf{u}_1,\mathbf{u}_2,\ldots]}$ are the basis vectors for the associated Graph Fourier Transform (GFT). 
The GFT coefficients $\mathbf{\tilde{f}}$ are computed as the scalar product between the graph signal $\mathbf{f}$ and each eigenvector:
\begin{equation}
    \mathbf{\tilde{f}} = \mathbf{U}^\top \mathbf{f}\,.
\end{equation}

Since the eigenvalues are real, they offer a natural method for ordering the GFT basis vectors according to frequency, reflecting the variations of their values across the graph. The eigenvalue/eigenvector pairs are successive optimizers of the Rayleigh quotient, where the $j$-th pair  
$(\lambda_j, \mathbf{u}_j)$ solves:
\begin{equation}\label{eq:quadratic_form_2}
    \mathbf{u}_j = \argminA_{\mathbf{z}^\top \mathbf{u}_{j'}=0,\,j'=0,\ldots,j-1} \frac{\mathbf{z}^\top \mathbf{L}\, \mathbf{z}}{\mathbf{z}^\top \mathbf{z}}
\end{equation}
and $\lambda_j = \mathbf{u}_j^\top \mathbf{L}\, \mathbf{u}_j$. Consequently, for the variation defined by the Laplacian quadratic form in \eqref{eq:quadratic_form}, the GFT yields an orthogonal basis with progressively increasing variation.

\begin{figure}[t]
\centering
	{\includegraphics[width=0.82\columnwidth]{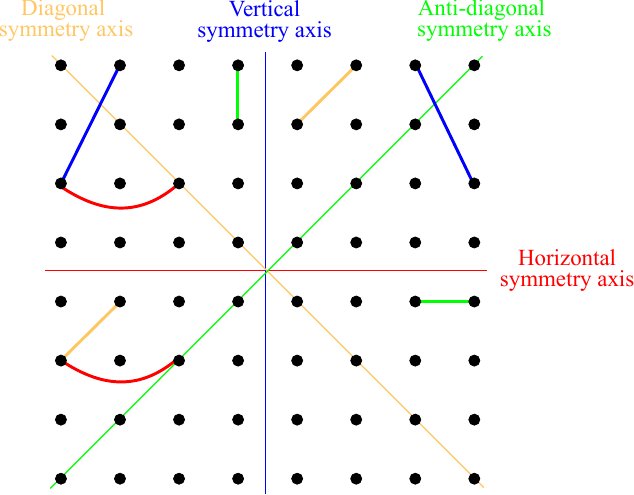}}
\caption{Grid axes and their corresponding edge symmetries for a 2DG. The vertical, diagonal, horizontal, and anti-diagonal axes are highlighted in blue, orange, red, and green, respectively. Example pairs of additional edges that would satisfy the ES property shown in parenthesis for the associated axis are drawn in matching colors.}
\label{fig:symmetries_on_grid}
\end{figure}

\subsection{Symmetry in 2D graphs}
\label{subsec:sym_graphs}
\noindent

As a starting point to build graphs representing $N{\times}N$ pixel blocks, let us consider a set of vertices $\mathcal{V}$ arranged in a \textit{2D grid} (2DG), where each of the $N_V=N^2$ vertices corresponds to one position on the $N{\times}N$ grid. 
Thus, we can arrange these nodes on a plane with integer coordinates $x$ and $y$ respectively\footnote{\, Herein $x$ is the row index increasing north to south and $y$ is the column index increasing west to east.}, in the range $1,\ldots, N$.
Connecting the nodes of a 2DG with edges leads to the following graph definition, 
which is the starting point for all our graph constructions:

\begin{definition}[2D grid graph]
\label{def:2DGG}
A \textit{2D grid graph} (2DGG) is formed from a 2DG by connecting every pair of horizontally and vertically adjacent grid vertices (i.e., at a distance of 1) with edges of weight 1. A 2DGG has $2N(N-1)$ edges.
\end{definition}
\noindent
Note that the 2DGG is associated with a Laplacian matrix whose eigenvectors correspond to the DCT basis \cite{strang1999discrete}.
%

The following definitions are needed to introduce 2DGG extensions with symmetry properties. 

\begin{definition}[Grid axes]
\label{def:grid_axis}
The four axes passing through the center of a 2DG at $0^\circ$, $45^\circ$, $90^\circ$, and $135^\circ$ angles are referred to as \textit{grid axes}, corresponding to the dashed lines in  Fig.~\ref{fig:symmetries_on_grid}.
\end{definition}

\begin{property}[Edge symmetry]
\label{def:ES}
A graph obtained by adding edges to the 2DGG has \textit{edge symmetry} (ES) if, for each pair of nodes connected by an edge, the corresponding pair of nodes in the mirrored position on the grid, relative to (at least) one of the grid axes, is also connected by an edge with the same weight. 
Specifically, the possible types of ES are (see  Fig.~\ref{fig:symmetries_on_grid}):
\begin{itemize}
    \item Left-right (LR)
    symmetry: 
    for any edge connecting two nodes at positions $(x_1, y_1)$ and $(x_2, y_2)$, there is a corresponding edge connecting the nodes at positions $(x_1, N - y_1+1)$ and $(x_2, N - y_2+1)$.
    \item Diagonal (D)
    symmetry:  
     for any edge connecting two nodes at positions $(x_1, y_1)$ and $(x_2, y_2)$, there is a corresponding edge connecting the nodes at positions $(y_1, x_1)$ and $(y_2, x_2)$.
    \item Up-Down (UD) symmetry: 
     for any edge connecting two nodes at positions $(x_1, y_1)$ and $(x_2, y_2)$, there is a corresponding edge connecting the nodes at positions $(N-x_1+1, y_1)$ and $(N-x_2+1, y_2)$.
    \item Anti-Diagonal (AD) symmetry: 
     for any edge connecting two nodes at positions $(x_1, y_1)$ and $(x_2, y_2)$, there is a corresponding edge connecting the nodes at positions $(N-y_1+1, N - x_1+1)$ and $(N-y_2+1, N - x_2+1)$.
\end{itemize}
\end{property}

\begin{definition}[Edge symmetric graphs]
\label{def:ESG}
A graph having an ES property of any type is an \textit{edge symmetric graph} (ESG).
\end{definition}

\begin{remark}
The 2DGG with no additional edges is an ESG that has all four ES types.
\end{remark}

Notably, ESGs have symmetric eigenvectors (Prop.~\ref{prop:sym_eig}) and these result in a faster implementation (Lemma~\ref{lem:fast}): 
\begin{property}[]
\label{prop:sym_eig}
In an ESG, there always exists a node ordering such that the associated Laplacian matrix is bisymmetric and its eigendecomposition leads to symmetric eigenvectors, that is, $\mathbf{u}_j(i)=\mathbf{u}_j(N+1-i),\ i,j=1,\ldots,N$ \cite{KSLU_TSP19, Cantoni1976eigenvalues}.
\end{property}
\begin{lemma}
\label{lem:fast}
Given Prop.~\ref{prop:sym_eig}, butterfly-based fast implementations can be used to compute the GFT of an ESG \cite{KSLU_TSP19}.
\end{lemma}

We now define an additional symmetry property, which, unlike ES (Def.~\ref{def:ES}), is a property of connected \textit{nodes}.

\begin{definition}[Reflection axis]
A \textit{reflection axis} $s$ is a line on the $(x,y)$ plane that traverses the 2DG, described by either $s:y=mx+q$ for some $m$ and $q$, or $s:x=a$ for some $a$, and which has at least one pair of nodes symmetric with respect to it\footnote{\,E.g., previously defined grid main axes can also be used as reflection axes.}.
\end{definition}
\begin{property}[Node symmetry]
\label{def:NS}
A graph constructed on a 2DG has \textit{node symmetry} (NS) with respect to a reflection axis $s$ if every pair of nodes symmetric relative to $s$ is connected by an edge.
\end{property}

\begin{figure}[t]
\centering
	{\includegraphics[width=0.7\columnwidth]{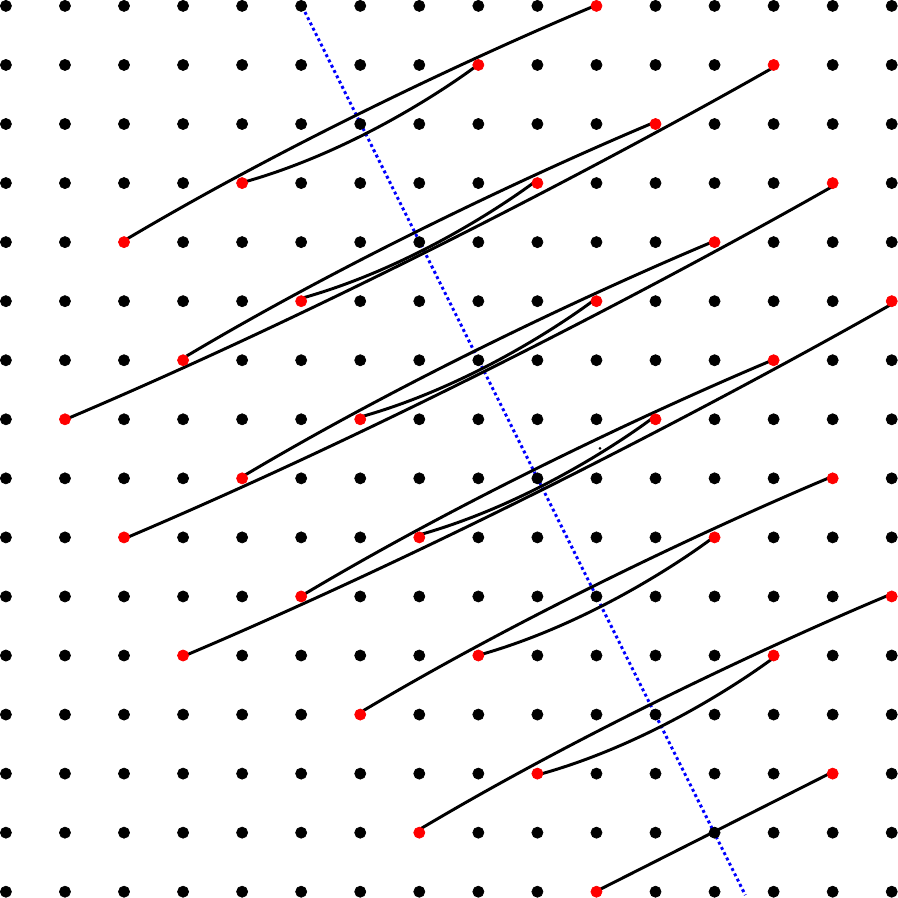}}
\caption{Example of an NS with respect to the dashed blue reflection axis. The red nodes for $\mathcal{V}_s$ and the black connections for $\mathcal{E}_s$ indicate the associated \textit{support}. More edges may be present without influencing the NS.}
\label{fig:gen_ref_axis}
\end{figure}

\begin{definition}[Support]
\label{def:support}
For a given graph with the NS property, the \textit{support} associated with $s$ is a subset $(\mathcal{V}_s,\mathcal{E}_s) \subset (\mathcal{V},\mathcal{E})$ that contains all the connected nodes symmetric with respect to $s$, as well as the corresponding edges that connect them. Refer to Fig.~\ref{fig:gen_ref_axis} for an example. 
\end{definition}


\section{Variable-Size Symmetry-Based GFTs}
\label{sec:sbgft}

Using the definitions of Sec.~\ref{sec:preliminaries} we propose a set of graphs.

\begin{figure*}[tbp]
\centering
\subfloat[][Horizontal reflection axes.]
	 {\includegraphics[scale=.65]{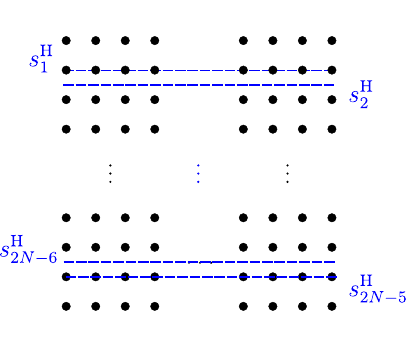}\label{fig:sym0}}\
\subfloat[][Anti-diagonal reflection axes.]
	 {\includegraphics[scale=.65]{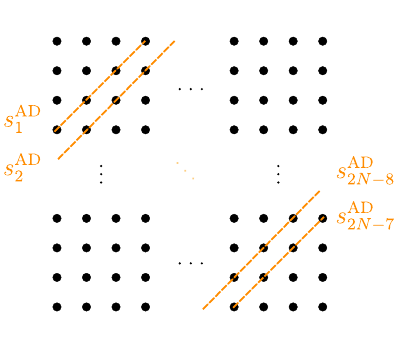}\label{fig:sym45}}\
\subfloat[][Vertical reflection axes.]
	 {\includegraphics[scale=.65]{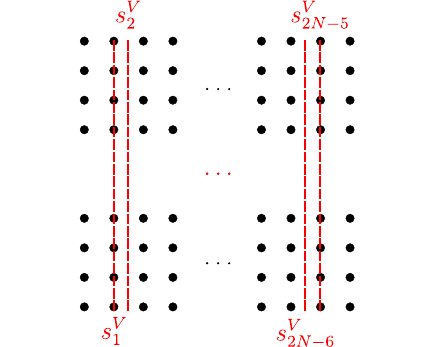}\label{fig:sym90}}\
\subfloat[][Diagonal reflection axes.]
	 {\includegraphics[scale=.65]{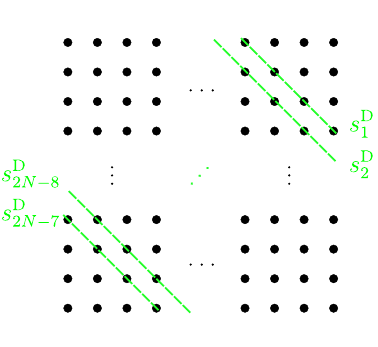}\label{fig:sym135}}
\caption{Position of all the reflection axes for a $N{\times}N$ grid. The reflection axes in each direction lead to SBGs with ES with respect to the main grid axis in the perpendicular direction, as shown by the maching color in Fig.~\ref{fig:symmetries_on_grid}.}
\label{fig:graphsymmetries}

\end{figure*}

\subsection{Design}
\label{subsec:design}

We extend the SBGFTs, originally designed for $8{ \times}8$ blocks in \cite{Gnutti_icip19}, to support blocks of size $N{\times}N$. While our generalization is also valid for odd values of $N$, we focus on even $N\geq4$ since partitioning in all image and video codecs is traditionally performed on even-sized blocks. SBGFTs are derived from the following set of graphs:

\begin{definition}[Symmetry-based graphs]
\label{def:SBG}
A \textit{Symmetry-based graph} (SBG) is built starting from the 2DGG and adding weight-$1$ node connections to induce an NS property with respect to a reflection axis directed in one of the four main directions, which is thus also \textbf{perpendicular} to a grid axis.
\end{definition}

\begin{remark}
The 2DGG without additional edges already exhibits the NS property for the axes $x = 1.5$, $x = N {-} 0.5$, $y = 1.5$, and $y = N {-} 0.5$. For example, the support for $x = N {-} 0.5$ contains the nodes in the last and second-to-last rows and the $N$ vertical edges connecting them. However, we do not consider the 2DGG as an SBG since no additional node connections on top of the grid are present.
\label{rem:2DGGrefl}\end{remark}

To build SBGs, the \textit{possible horizontal (vertical) reflection axes} $s_i^{\text{H}}:x=q_h$ ($s_i^{\text{V}}:y=q_v$) are such that $q_h$ ($q_v$) should assume one of the integer or half-integer values between $2$ and $N-1$, as in Figs.~\ref{fig:sym0} and \ref{fig:sym90}:
\begin{align}
    s^{\text{H}}_k & :x=\frac{k+3}{2}, & k & =\{1,\ldots,2N-5\}\label{eq:axesLR}\,,\\
    s^{\text{V}}_k & :y=\frac{k+3}{2}, & k & =\{1,\ldots,2N-5\}\label{eq:axesUD}\,.
\end{align}
\noindent
For example, the SBG associated with $s^{\text{H}}_1:x=2$ is the 2DGG with added connections between the nodes in $(1,j)$ and $(3,j)$ for $j=1,\ldots,N$. 

Algorithm~\ref{alg:edgesUD-LR} summarizes SBGs construction from horizontal and vertical reflection axes. The symbol $\leftrightarrow$ denotes a connection between specular nodes to be added to the graph.

Next, the \textit{possible diagonal (anti-diagonal) reflection axes} $s_i^{\text{D}}:y=x+q_d$ ($s_i^{\text{AD}}:y=-x+q_{ad}$) are such that $q_d$ ($q_{ad}$) lead the axes to pass through the nodes, thus must be integer:
%
\begin{align}
    s^{\text{D}}_k&:y=x-(N-3)+k, & k &=\{1,\ldots,2N-7\}\label{eq:axesD}\,,\\
    s^{\text{AD}}_k&:y=-x+4+k, & k &=\{1,\ldots,2N-7\}\label{eq:axesAD}\,.
\end{align}
Note that these equations leave out the axes passing through the last $3$ nodes in the grid corners, e.g., the north-east corner nodes $(1,N{-}2)$, $(1,N{-}1)$, and $(1,N)$ for the diagonal direction, so that the support of the NS property of each SBG has a minimum of $6$ added edges (see Figs.~\ref{fig:sym45} and \ref{fig:sym135}).

Algorithm~\ref{alg:edgesD-AD} shows how SBGs are built from diagonal and anti-diagonal reflection axes. To shorten the pseudo-code,
each iteration of $n$ can generate two graphs, since each of its $N{-}3$ values identifies a pair of reflection axes in either the diagonal or anti-diagonal direction. As the first iteration for $n=1$ ($k={N{-}3}$) corresponds to the main grid axes, it generates the same graph for both instructions, bringing the total number of built graphs for each symmetry to $2N{-}7$.

Therefore, the final SBG set is composed of $8N{-}24$ elements, one for each valid reflection axis. 
The GFTs associated with this set of graphs are called Symmetry-Based Graph Fourier Transforms (SBGFTs).

\begin{algorithm}[t]
\caption{SBG building for $s^H$ and $s^V$}
\label{alg:edgesUD-LR}
\begin{algorithmic}[1]
\REQUIRE Block size $N$
    \FOR{$k \gets 1$ to $2N-5$}
        \FOR{$j \gets 1$ to $N$}
            \FOR{$i \gets 1$ to $\min\left(\lfloor\frac{k+2}{2}\rfloor),N-\lfloor\frac{k+3}{2}\rfloor\right)$}
                \STATE $a \gets \frac{k+3}{2}-i+\frac{k+3}{2} \bmod 1$
                \STATE $b \gets \frac{k+3}{2}+i-\frac{k+3}{2} \bmod 1$
                \STATE LR symmetries: \STATE $\ (j,a)\leftrightarrow(j,b)$
                \STATE UD symmetries:
                \STATE $\ (a,j)\leftrightarrow(b,j)$
            \ENDFOR
        \ENDFOR
    \ENDFOR
   
\end{algorithmic}
\end{algorithm}

\begin{algorithm}[t]
\caption{SBG building for $s^D$ and $s^{AD}$}
\label{alg:edgesD-AD}
\begin{algorithmic}[1]
\REQUIRE Block size $N$
    \FOR{$n{\,}^* \gets 1$ to $N-3$}
        \FOR{$j \gets 1$ to $N-n$}
            \FOR{$i \gets n+j$ to $N$}
                \STATE $a \gets j+(n-1)$
                \STATE $b \gets i-(n-1)$
                \STATE AD symmetries${\,}^{**}$:
                \STATE $\ $ $(i,j)\leftrightarrow(a,b)$
                \STATE $\ $ $(N{+}1{-}j,N{+}1{-}i)\leftrightarrow(N{+}1{-}b,N{+}1{-}a)$
                \STATE D symmetries${\,}^{**}$:
                \STATE $\ $ $(a,N+1-b)\leftrightarrow(i,N+1-j)$
                \STATE $\ $ $(b,N+1-a)\leftrightarrow(j,N+1-i)$
            \ENDFOR
        \ENDFOR
    \ENDFOR
   
\end{algorithmic}
${\ }^*$\, $n=|k-(N-3)|+1$, with $k$ as in (\ref{eq:axesD}) and (\ref{eq:axesAD}).\\
${\ }^{**}\,$For $n{=}1$, instructions 7--8 and 10--11 give the same edges.
\end{algorithm}

\subsection{Properties}
\label{subsec:prop}

\begin{figure}[t]
\centering
\subfloat[][LR edge symmetry ($s_2^{\text{H}}$).]
	 {\includegraphics[height=4cm]{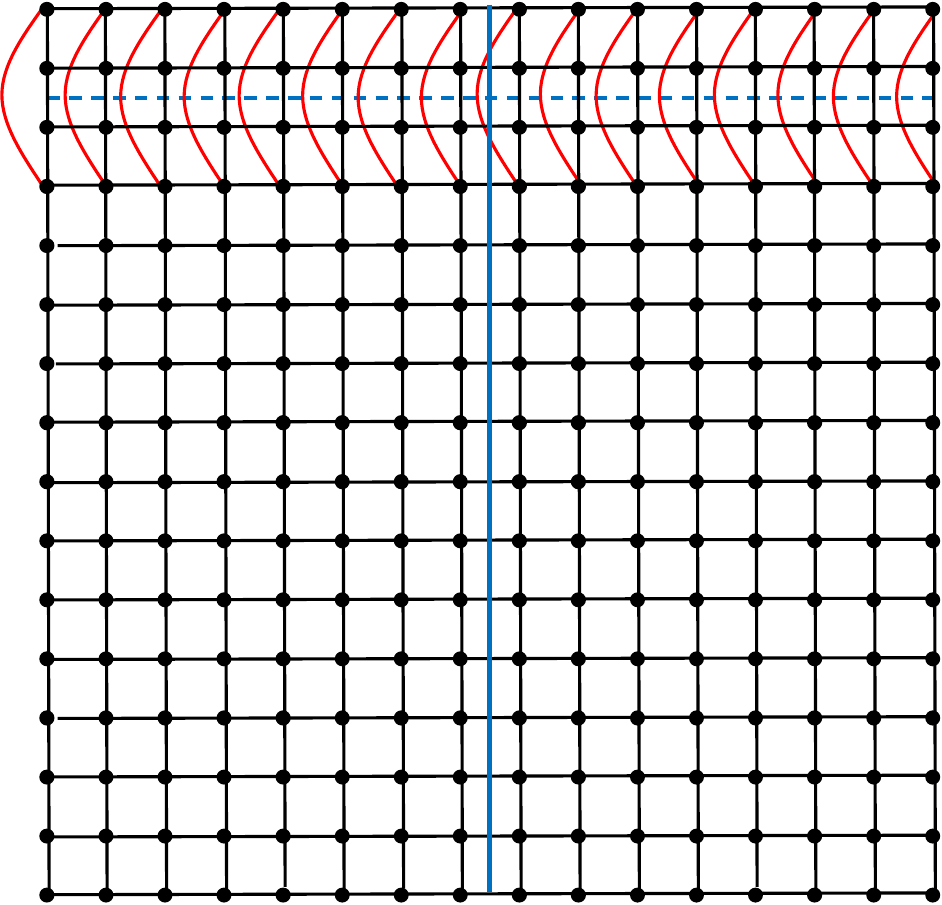}\label{fig:lr}}\quad
\subfloat[][D edge symmetry ($s_2^{\text{AD}}$).]
	 {\includegraphics[height=4cm]{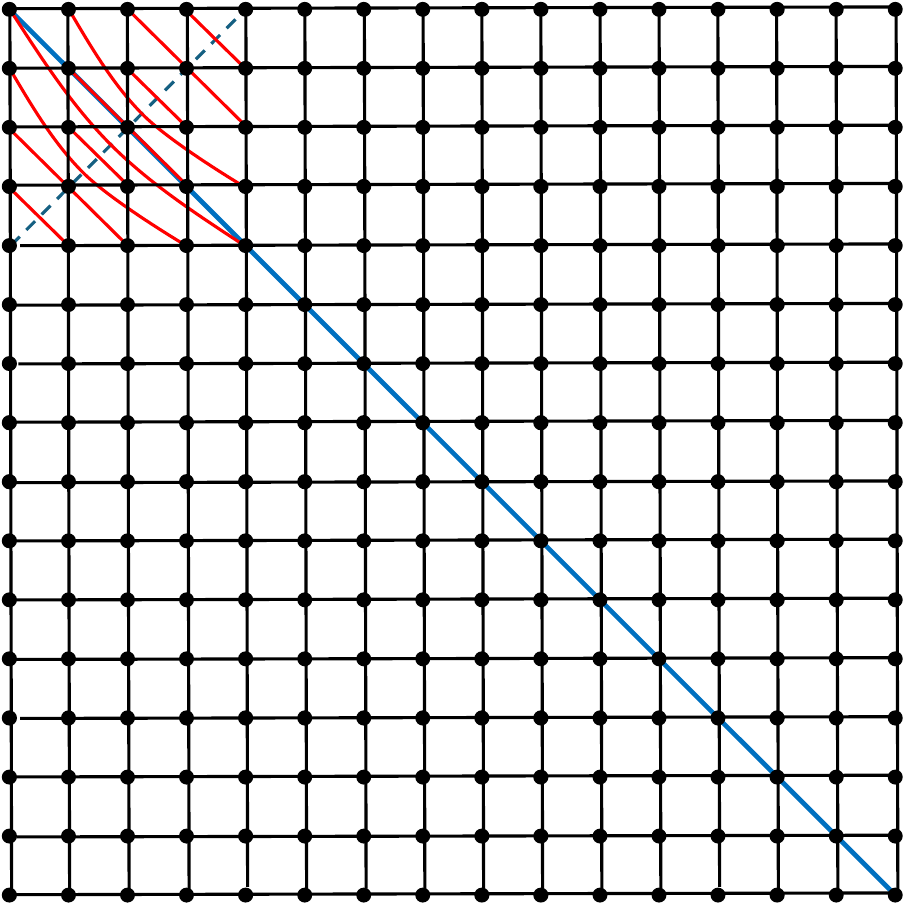}\label{fig:diag}}\\
\hspace{0.1cm}
\subfloat[][UD edge symmetry ($s_2^{\text{V}}$).]
	 {\includegraphics[height=4.15cm]{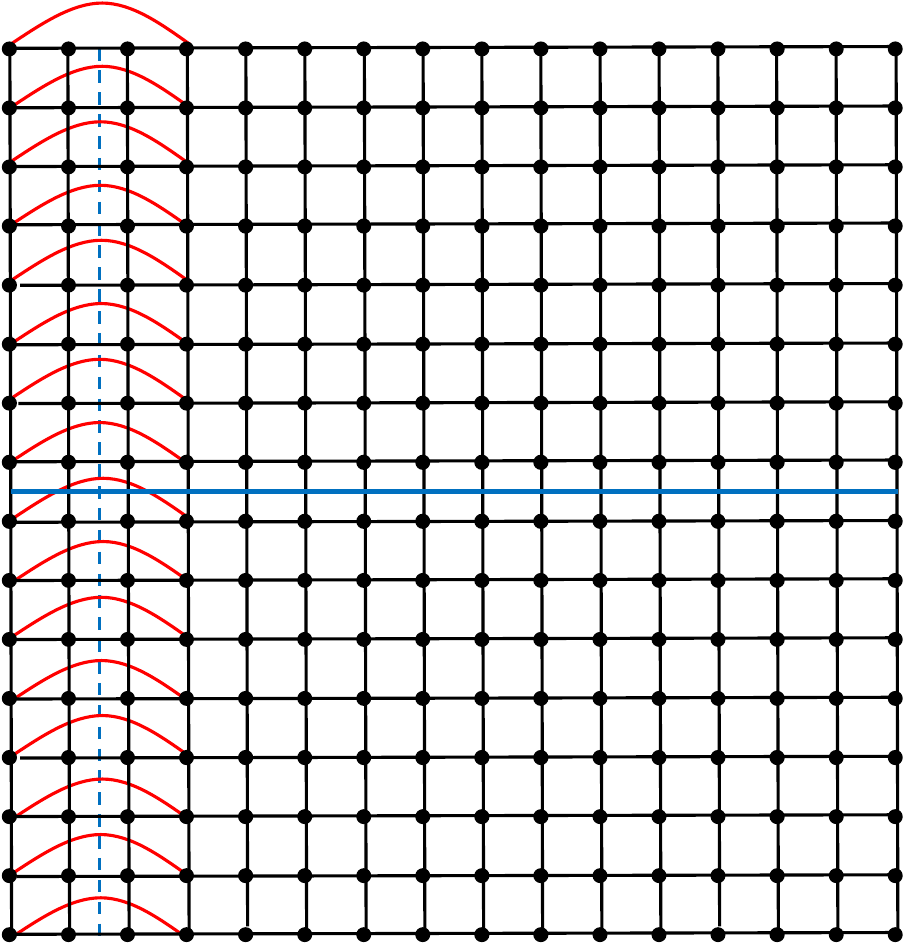}\label{fig:ud}}\quad
\subfloat[][AD edge symmetry ($s_2^{\text{D}}$).]
	 {\includegraphics[height=4cm]{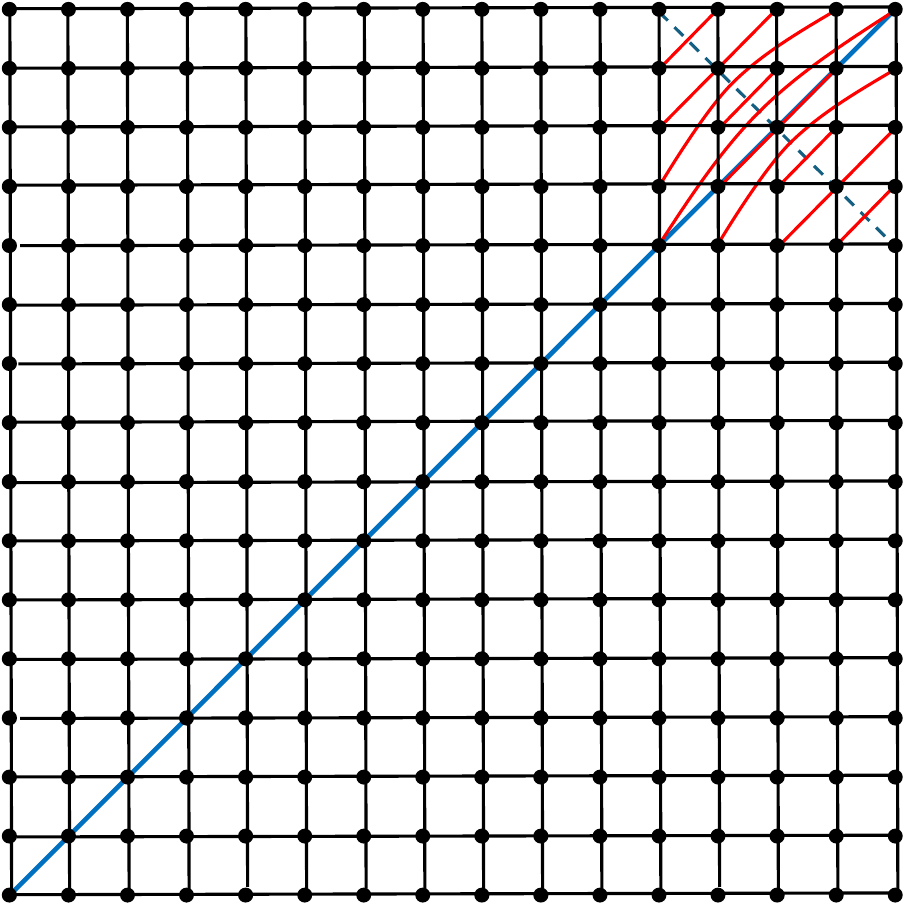}\label{fig:anti-diag}}
\caption{Example of SBGs constructed on a $16{\times}16$ 2DGG, putting $k=2$ in \eqref{eq:axesLR}--\eqref{eq:axesAD}. Each graph has ES with respect to one of the grid axes highlighted by the continuous blue lines. Black edges are the baseline grid edges, while red edges are added edges that induce the NS property with respect to the reflection axis drawn as a dashed blue line.}
\label{fig:16x16_examples}
\end{figure}

While non-separable transforms generally have higher computational complexity than separable transforms, the symmetry properties of our proposed non-separable SBGFTs lead to faster implementation. Indeed, SBGs and ESGs are related:

\begin{property}[]
A SBG, built to possess NS with respect to a reflection axis orthogonal to a grid axis, is also an ESG. Specifically, NS with reflection axes $x = q_h$ (horizontal), $y = q_v$ (vertical), $y = x + q_d$ (diagonal), and $y = -x + q_{ad}$ (anti-diagonal) induce the LR, UD, AD, and D types of ESGs, respectively (see Fig.~\ref{fig:16x16_examples}).
\end{property}

Exploiting the eigenvector symmetry for ESG (Property \ref{prop:sym_eig}), using the results in \cite{KSLU_TSP19}, the number of multiplications to compute each type of ES can be reduced, as shown in Table~\ref{tab:speedup}. For context, non-symmetric transforms need $2N^3$ multiplications when separable, and $N^4$ multiplications otherwise. The $4$ graphs obtained using one of the main grid axes as the reflection axis (last two rows of Table~\ref{tab:speedup}) enjoy a double ES and their GFTs can be computed even more efficiently.In addition to this type of symmetry present in the SBG eigenvectors, another form of symmetry emerges. In fact, from \eqref{eq:quadratic_form}, when an edge is added to the graph, a higher variation is associated with signals having different values in the newly connected nodes. Thus, from \eqref{eq:quadratic_form_2}, those newly connected nodes will tend to have similar values in low-frequency (low-variation) eigenvectors. This implies that the low-frequency eigenvectors of an SBG will display local symmetry properties, i.e., connected nodes symmetric relative to the reflection axis will have similar values.

To quantitatively verify this eigenvector symmetry, we first define, for a given symmetry axis $s$, the symmetry ratio of a graph signal. Multiple symmetry ratios can be computed for graphs enjoying multiple symmetry axes.

\begin{definition}[Mirroring operation with respect to $s$]
\label{def:sym_ratio}
Given a reflection axis $s$, let $\mathbf{f}^{(s)}$ be the $|\mathcal{V}_s|$-long column vector representing the portion of a graph signal defined on its support $\mathcal{V}_s$ (Def.~\ref{def:support}). The mirroring operation $m\left( \mathbf{f}^{(s)},s \right )$ of $\mathbf{f}^{(s)}$ with respect to $s$ is defined as exchanging each $\mathbf{f}^{(s)}$ sample across nodes connected by the edges in $\mathcal{E}_s$, that is, swapping the graph signal values on each node in the support with those in specular positions with respect to $s$.
\end{definition}

\begin{definition}[Symmetry ratio on support $\mathcal{V}_s$]
We define the \textit{symmetry ratio} $S_s$ of $\mathbf{f}^{(s)}$ with respect to $s$ as:
\begin{equation}
    S_s = \frac{{\mathbf{f}^{(s)}}^\top \cdot m\left( \mathbf{f}^{(s)},s \right )}{{\mathbf{f}^{(s)}}^\top \cdot \mathbf{f}^{(s)}}\,.
\end{equation}
The symmetry ratio has values in $[-1,1]$, with $1$ and $-1$ meaning perfect symmetry and anti-symmetry, respectively.
\end{definition}

\begin{figure}[t]
\centering
	\subfloat[$|S_s|$ for $s_1^{\text{H}},\ldots,s_5^{\text{H}}$.]{\includegraphics[width=0.48\columnwidth]{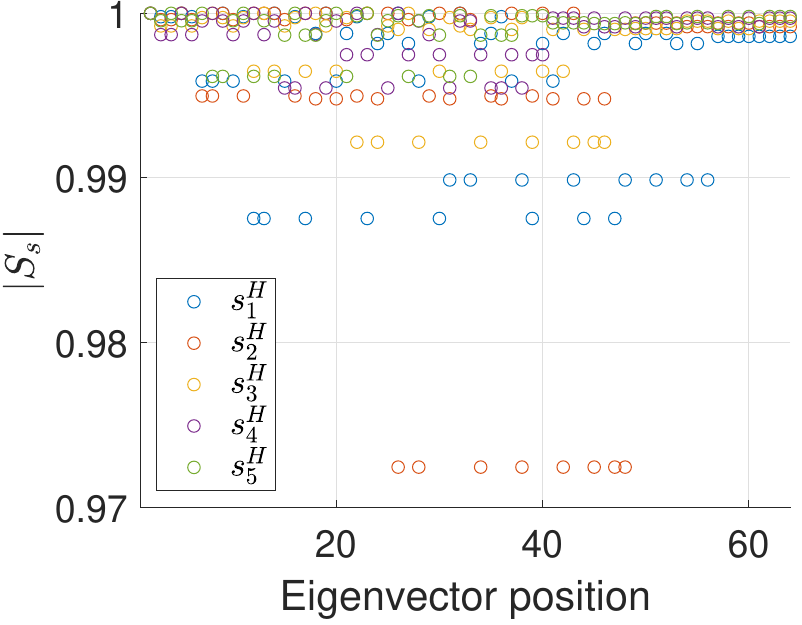}}\quad
    \subfloat[$|S_s|$ for $s_1^{\text{V}},\ldots,s_5^{\text{V}}$.]{\includegraphics[width=0.48\columnwidth]{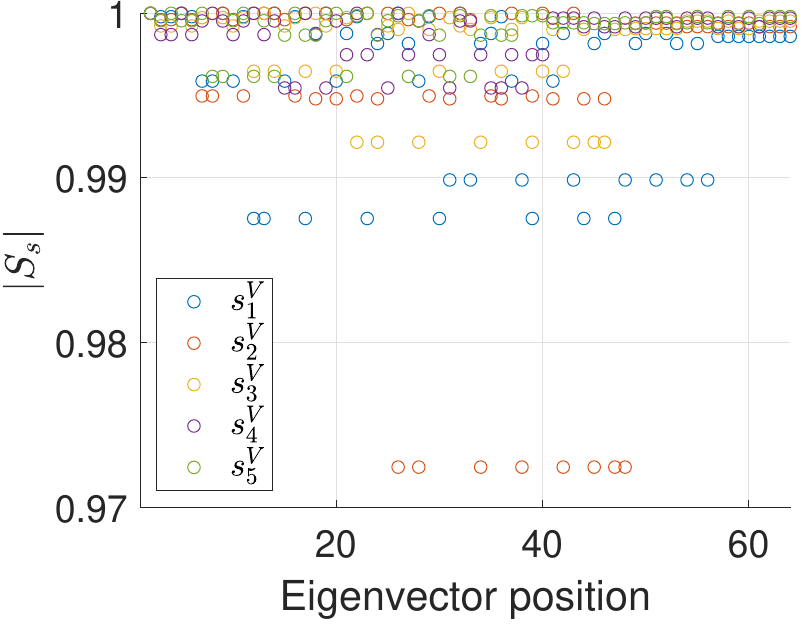}}\\
    \subfloat[$|S_s|$ for $s_1^{\text{D}},\ldots,s_4^{\text{D}}$.]{\includegraphics[width=0.48\columnwidth]{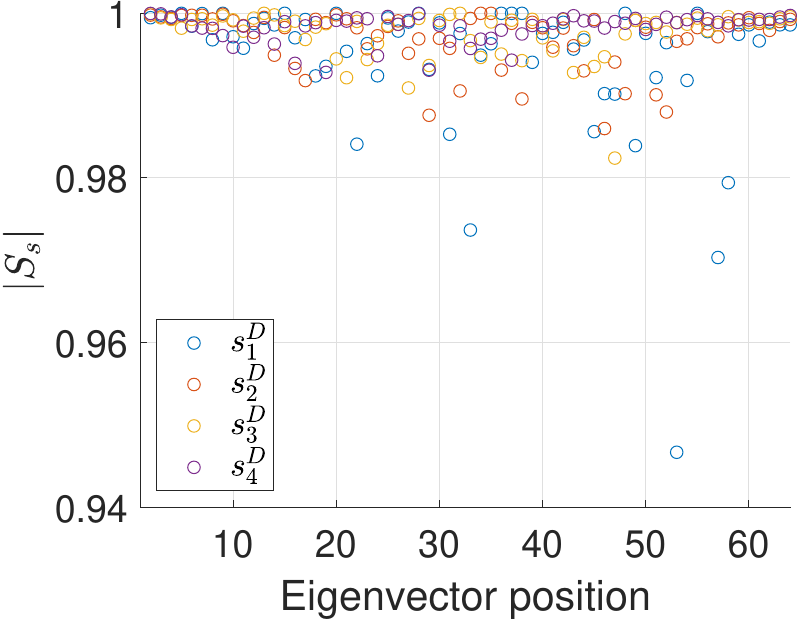}}\quad
    \subfloat[$|S_s|$ for $s_1^{\text{AD}},\ldots,s_4^{\text{{AD}}}$.]{\includegraphics[width=0.48\columnwidth]{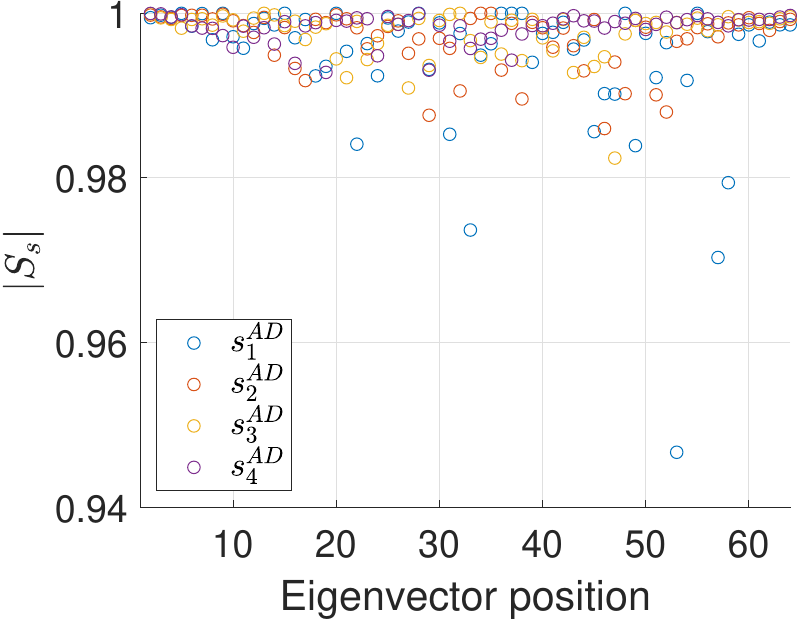}}
\caption{Symmetry ratios of all eigenvectors for $8{\times} 8$ SBGs built with reflection axes different than main grid axes. Only the first half of the axes are depicted since the other half enjoys the same symmetry ratio. The SBGs are grouped by the reflection axes direction: (a) horizontal, (b) vertical, (c) diagonal, and (d) anti-diagonal.}
\label{fig:scatter}
\end{figure}

Based on our previous discussion, nodes that become connected tend to have similar values in low-frequency eigenvectors. Thus, in our SBGs set, we expect that eigenvector values localized to the nodes in $\mathcal{V}_s$, say $\mathbf{u}_j^{(s)}$ for the $j$-th eigenvector, will enjoy higher symmetry. To show this, Fig.~\ref{fig:scatter} presents the symmetry ratios for all the eigenvectors associated with each SBG as built by reflection axes different from the main grid axes for the case $N=8$. As a matter of fact, for these axes, the minimum symmetry ratio $S_s$ is $0.94$. Instead, for the main grid axes, we can observe the following:

\begin{table}[t]
\caption{Number of multiplications to compute the speeded up GFT for the proposed $N{\times}N$ ESGs.}\label{tab:speedup}
\centering\setcellgapes{4pt}\makegapedcells \renewcommand\theadfont{\normalsize\bfseries}
\centering
\begin{tabular}{l|c|c}
{\bf ES type} & {\bf \# of graphs} & {\bf \# of multiplications}\\
\addlinespace[2pt]\toprule
UD & $2N-4$ &$N^4/2$\\
\midrule
LR & $2N-4$ &$N^4/2$\\
\midrule
D & $2N-6$ &$N^2(N^2+1)/2$\\
\midrule
AD& $2N-6$ &$N^2(N^2+1)/2$\\
\midrule
UD and LR & 2 &$N^4/4\, \stackrel{N<9}{\sim}\, 2N^3$\\
\midrule
D and AD & 2 &$N^2(N^2+1)/4\, \stackrel{N<9}{\sim}\, 2N^3$\\
\bottomrule
\end{tabular}
\end{table}

\begin{remark}
\label{rem:perfsym}
For the $4$ SBGs obtained using one of the main grid axes as the reflection axis $s$, the eigenvectors $\mathbf{u}_j^{(s)}$ defined on the corresponding support $\mathcal{V}_s$ have $S_s = |1|$. Specifically, half of the eigenvectors associated with the lowest eigenvalues have $S_s = 1$, while the other half associated with the highest eigenvalues have $S_s = -1$.
\end{remark}

This is due to the fact that the Laplacian matrix associated to these $4$ SBGs is already bisymmetric without needing any node reordering as prescribed in Property \ref{prop:sym_eig}. So, the associated eigenvectors are symmetric in the sense described in that property, and this is the underlying reason for the associated GFT computational speedup already mentioned in Table~\ref{tab:speedup}.

In addition, following \cite[Th.~2 on p.~280]{Cantoni1976eigenvalues}, a pair of $N_V/2{\times}N_V/2$ matrices (recall that $N_V=N^2$) can be obtained from the Laplacian matrix to compute $N_V/2$-long eigenvectors, that are proportional to the first $N_V/2$ values of the starting Laplacian eigenvectors (the other $N_V/2$ values are obtained by flipping these values since the eigenvectors have to be symmetric).

For the aforementioned $4$ SBGs these $N_V/2{\times}N_V/2$ matrices contain bisymmetric blocks and can be written as:
\begin{equation}
\left[\begin{array}{cccc}
\mathbf{A}&\mathbf{JBJ}&\mathbf{JCJ}&\mathbf{0}\\
\mathbf{B}&\mathbf{JAJ}&\mathbf{0}&\mathbf{C}\\
\mathbf{C}&\mathbf{0}&\mathbf{D}&\mathbf{JEJ}\\
\mathbf{0}&\mathbf{JCJ}&\mathbf{E}&\mathbf{JDJ}
\end{array}\right]\,,
\end{equation}
where $\mathbf{J}$ is the reversal matrix and each block is $N_V/8{\times}N_V/8$. These block symmetries are reflected in the eigenvectors and eigenvalues computation, as they possess further symmetric structure, which is the reason for the validity of Remark \ref{rem:perfsym}. Therefore, when the eigenvectors are laid on these $4$ graphs as signals using standard column-wise node ordering, they have maximum symmetry ratio $S_s$, that is, the signal value on the nodes connected by the additional edges are always the same (or with the sign reversed).

\begin{figure*}[p]
\centering
\subfloat[][Symmetry ratio on $8 \! \times \! 8$ blocks.]
	 {\includegraphics[width = 0.48\columnwidth]{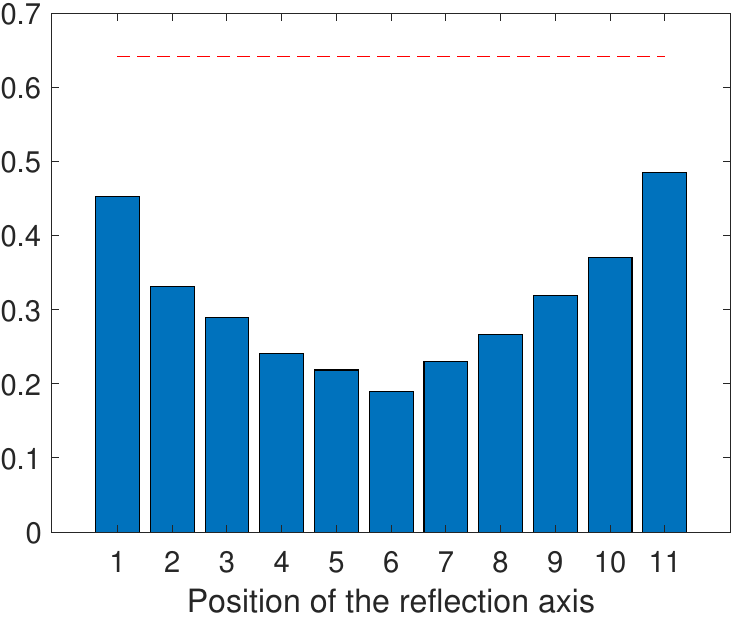}\quad
    \includegraphics[width = 0.48\columnwidth]{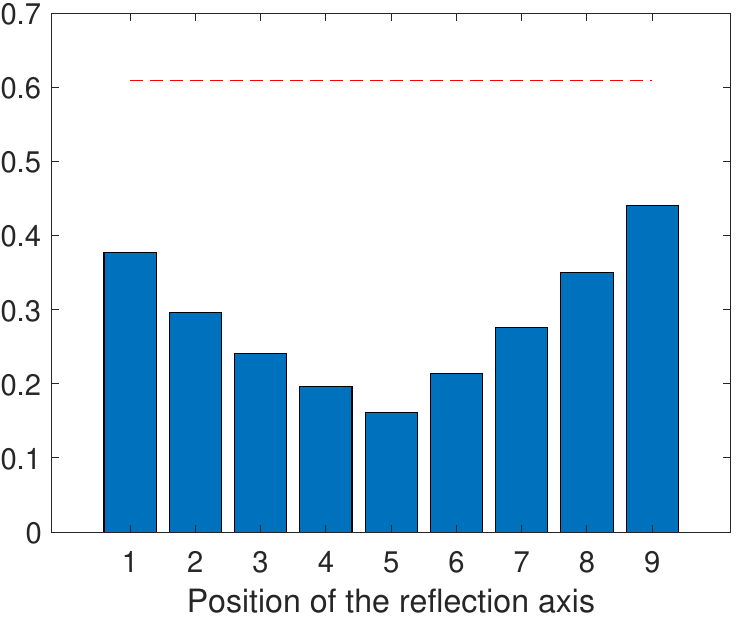}\quad
    \includegraphics[width = 0.48\columnwidth]{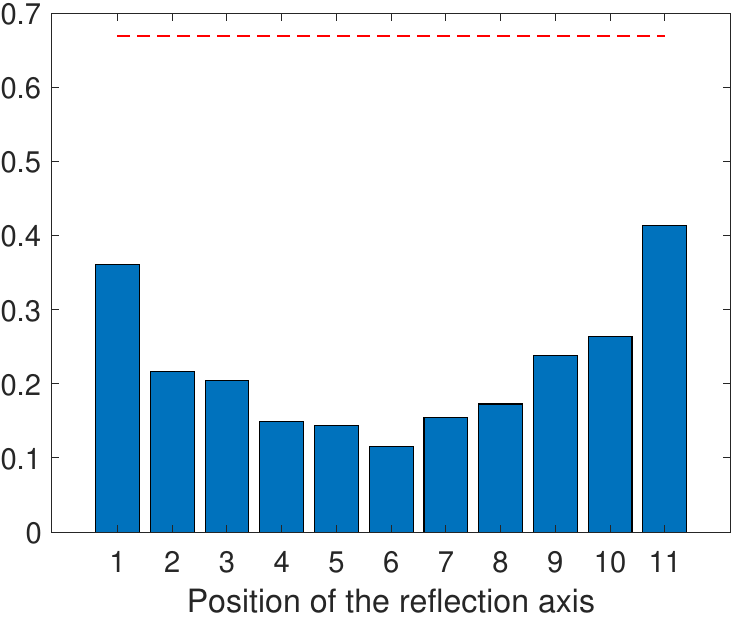}\quad
    \includegraphics[width = 0.48\columnwidth]{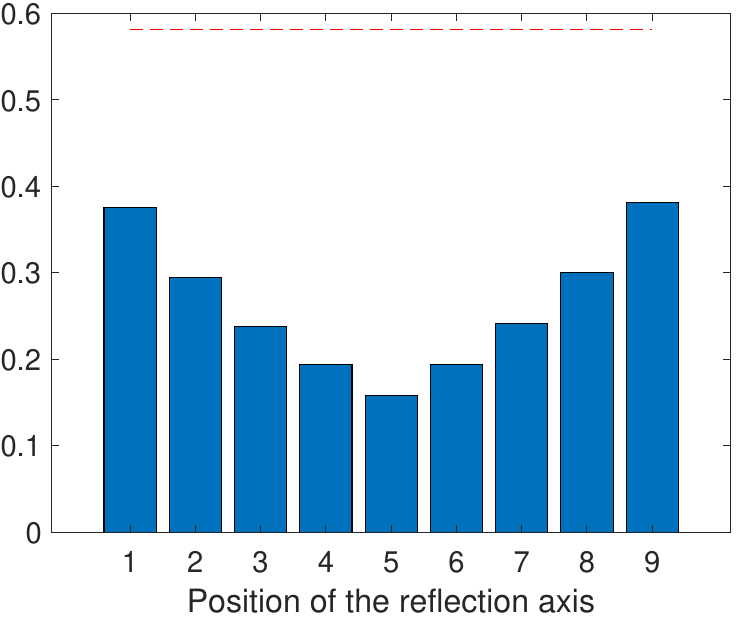}\label{fig:sym_in_residuals_8}}\\
\subfloat[][Symmetry ratio on $16 \! \times \! 16$ blocks.]
	 {\includegraphics[width = 0.48\columnwidth]{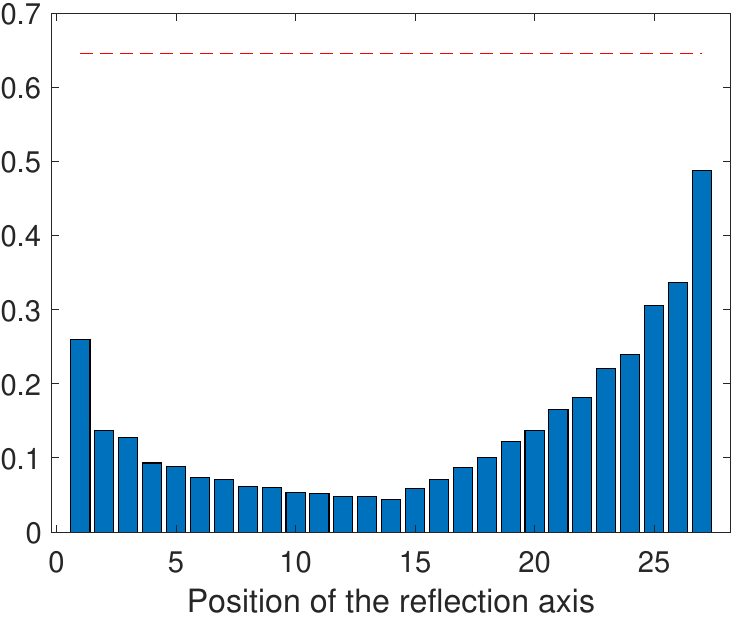}\quad
    \includegraphics[width = 0.48\columnwidth]{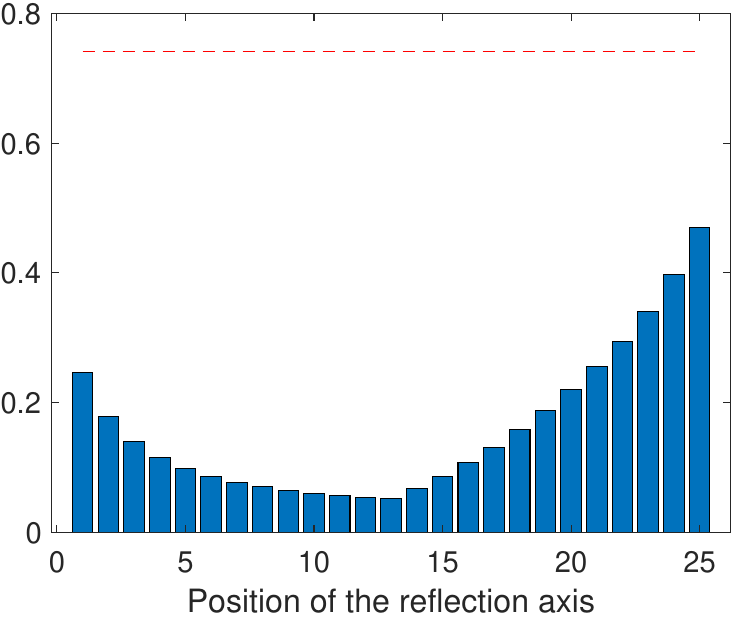}\quad
    \includegraphics[width = 0.48\columnwidth]{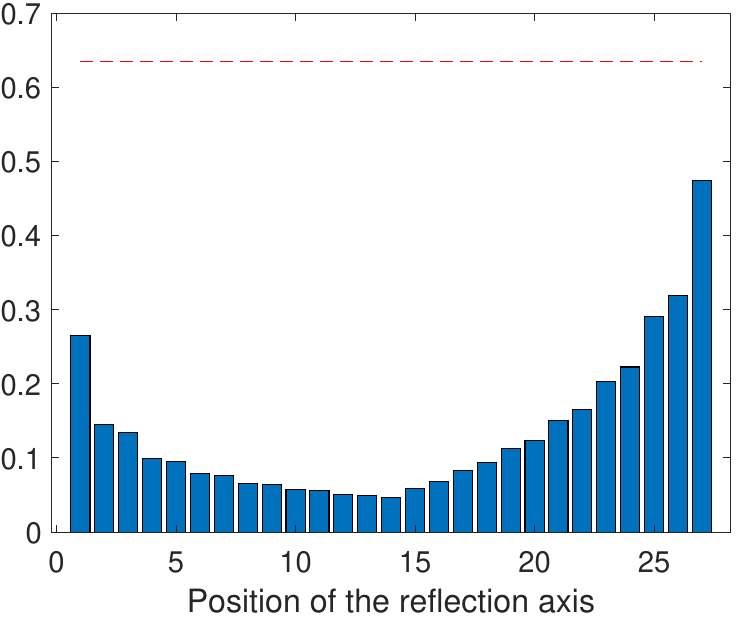}\quad
    \includegraphics[width = 0.48\columnwidth]{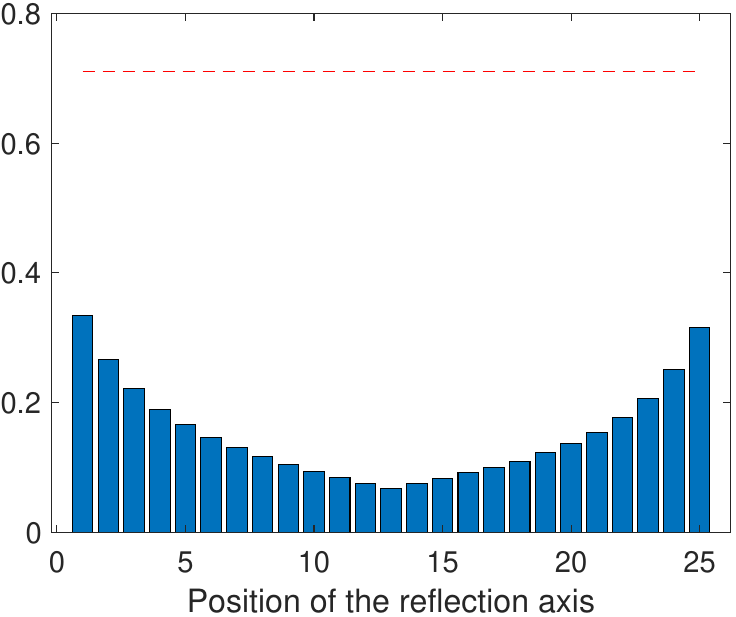}\label{fig:sym_in_residuals_16}}\\
    \subfloat[][Symmetry ratio on $32 \! \times \! 32$ blocks.]
	 {\includegraphics[width = 0.48\columnwidth]{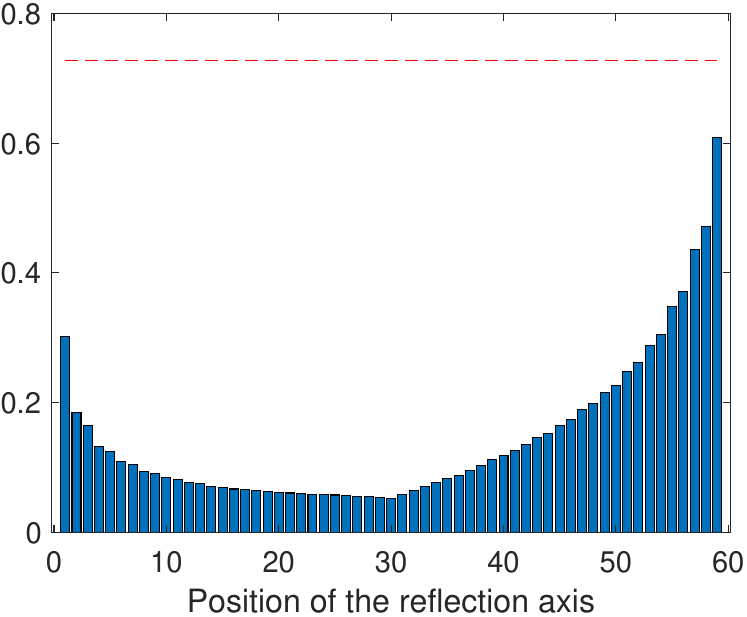}\quad
    \includegraphics[width = 0.48\columnwidth]{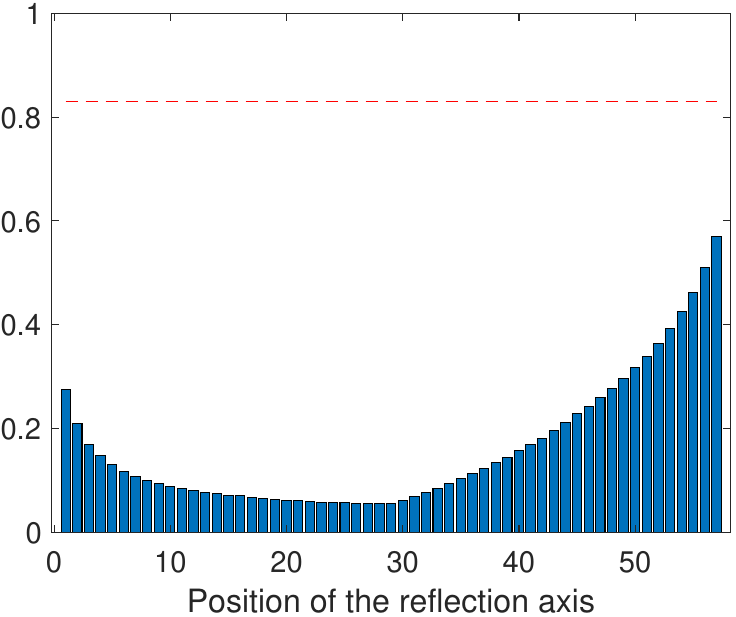}\quad
    \includegraphics[width = 0.48\columnwidth]{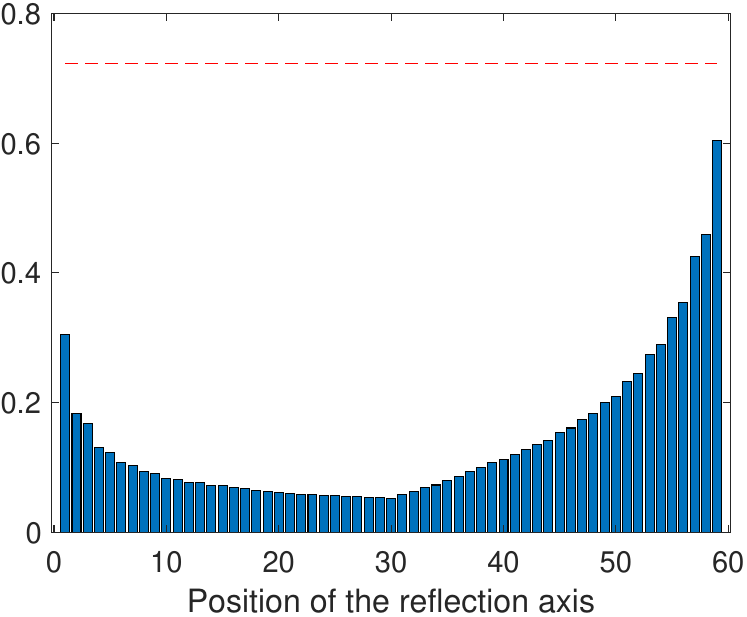}\quad
    \includegraphics[width = 0.48\columnwidth]{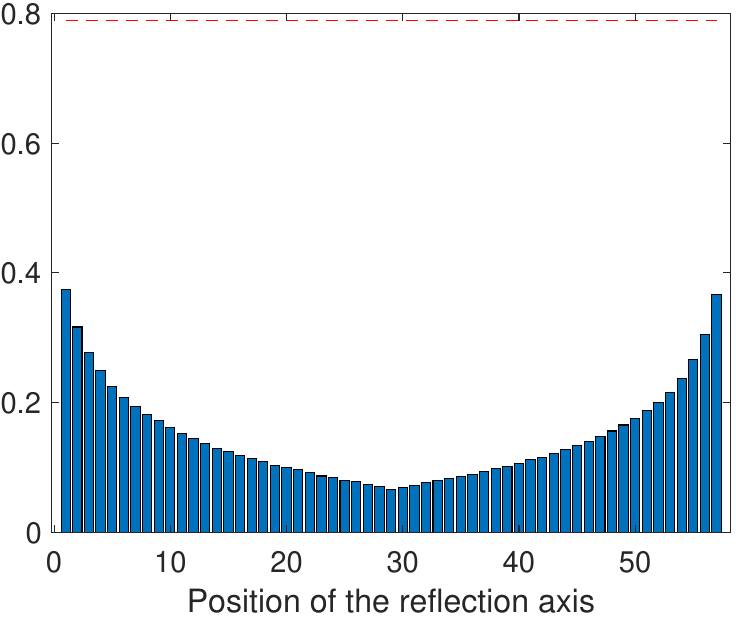}\label{fig:sym_in_residuals_32}}\\
\caption{Symmetry ratios in residual blocks of sizes (a) $8{ \times}8$, (b) $16{\times}16$, and (c) $32{\times}32$. The analyzed symmetries (LR, D, UD, and AD) are shown from left to right. Each bar represents the relative frequency of residual blocks with a symmetry ratio $S_s > 0.7$ for the respective reflection axis. The red dashed line indicates the percentage of residual blocks that achieve an $S_s$ greater than 0.7 in at least one symmetry.}
\label{fig:sym_in_residuals}
\end{figure*}
\begin{figure*}[p]
    \centering
    \includegraphics[width = 0.98\textwidth]{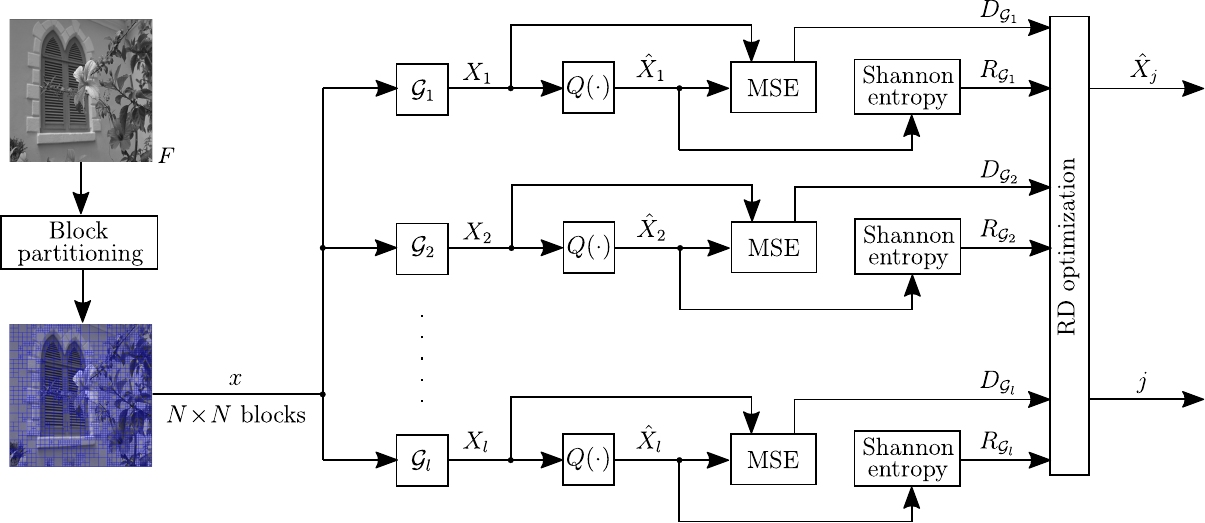}
    \caption{The multiple transform-based compression framework for evaluating the performance of the variable size SBGFTs. The set of SBGFTs $\mathcal{G}_{i},\ i=1,\ldots,l=8N-24$ is different for each block size $N$. The final arithmetic coding is not depicted. }
    \label{fig:overview_standard}
\end{figure*}

\subsection{Motivation: symmetry in residuals}
\label{subsec:sym_residuals}
\noindent

In this section, we provide evidence of how the local symmetry properties of the SBG eigenvectors contribute to the high efficiency of the resulting SBGFTs in representing residual data. To do so, we conduct a different analysis to show that the residual blocks derived from intra-prediction exhibit significant symmetry along the reflection axes perpendicular to the four grid axes. Since the eigenvectors' structure closely matches the residual blocks' inherent structure, SBGFTs are especially efficient in representing this type of data, as will be shown in Sec.~\ref{sec:exp}.

A preliminary analysis of symmetric content in residuals was presented in \cite{Lu-icassp-2017}, where the authors examined symmetry in $8{\times}8$ residual blocks following HEVC intra-prediction, using the four grid axes as reflection axes. Their findings showed that $36\%$ of the residual blocks displayed a high degree of symmetry. Building on this study, we have collected a large dataset of variable-sized residual blocks generated from VVC intra-prediction, including all prediction modes. Then, we have evaluated the symmetry ratio $S_s$ defined in Def.~\ref{def:sym_ratio} for each block related to the reflection axes specified in \eqref{eq:axesLR}--\eqref{eq:axesAD}.

Fig.~\ref{fig:sym_in_residuals} presents our findings for $8{ \times}8$, $16{\times}16$, and $32{\times}32$ residual blocks, which prove the high degree of local symmetry inherent in the residual data. For instance, examining the fourth distribution in Fig.~\ref{fig:sym_in_residuals_32}, we observe that approximately 10\% of the residual blocks have an $S_s$ greater than 0.7 along the anti-diagonal symmetry axis (center of the plot). Additionally, for more lateral sub-diagonals, $S_s$ increases, reaching nearly 40\%. Overall, 80\% of the blocks exhibit an $S_s \geq 0.7$ in at least one of the AD reflection axes.
These results are consistent across different block sizes and reflection axes, indicating that a significant number of residual blocks exhibit a high degree of symmetry with respect to the axes used to build the SBGs.


\section{Experimental results}
\label{sec:exp}
\noindent
This section describes several experiments conducted on pixel data and intra-prediction residuals. Various experimental configurations are tested to explore the benefits of variable-size SBGFTs, varying both the partitioning strategy and the transforms that can be applied for each block size. As in the rest of the paper, we only consider square-shaped blocks.


Before delving into the experiments in the pixel domain (Sec.~\ref{subsec:firstrun}) and residuals domain (Sec.~\ref{subsec:secondrun}), the framework for employing multiple transforms, such as the SBGFTs, using the RDOT scheme is illustrated in Sec.~\ref{subsec:framework}. Complexity issues are discussed in Sec.~\ref{subsec:complexity_analysis}, and, following the conclusions therein, a low complexity alternative to fully fledged variable size SBGFTs sets is proposed in Sec.~\ref{subsec:low_complexity}.

\subsection{Multiple transforms framework}
\label{subsec:framework}
\noindent
The variable size SBGFTs discussed in Sec.~\ref{sec:sbgft} are incorporated within a multiple transform compression scheme, similar to modern image and video compression standards. This integration aims to evaluate the potential advantages of employing the SBGFTs in a RDOT procedure \cite{zhao2011video} compared to using only traditional transforms such as the DCT and its variants, which serve as the primary transforms in VVC. 

As shown in Fig.~\ref{fig:overview_standard}, block partitioning subdivides the input visual data unit $F$ into square blocks with $N=2^n,\,n=2,\ldots,6$, that is, with sizes spanning from $4{\times}4$ to $64{\times}64$, as in both BPG and VVC. We postpone discussing the specifics of the partitioning strategy employed in our experiments to Secs.~\ref{subsec:firstrun} and \ref{subsec:secondrun}.

The optimal GFT $\mathcal{G}_{j}$ for each $N{\times}N$ block $\mathbf{x}$ is identified through the minimization of the Lagrangian cost function:
\begin{equation}\label{eq:lagrangian}
    j = \argmin_i D_{\mathcal{G}_{i}} + \lambda\, R_{\mathcal{G}_{i}}
\end{equation}
with the Lagrange multiplier $\lambda \ge 0$ determining the RD working point. $D_{\mathcal{G}_{i}}$ is obtained by computing the mean squared error (MSE) between the transform coefficients $\mathbf{X}_i$ and $\mathbf{\hat{X}}_i$, the quantized version of $\mathbf{X}_i$ using a given quantization parameter (QP). The rate $R_{\mathcal{G}_{i}}$ is approximated with the estimated Shannon entropy of $\mathbf{\hat{X}}_i$.

\newcommand{\colwid}{1.65cm}
\newcommand{\colwidd}{1.9cm}
\begin{table*}[t]
\caption{Summary of partitioning methods and employed transforms for the configurations used in our experiments.}
\centering
\begin{tabular}{lcP{\colwid}@{\ }P{\colwid}@{\ }P{\colwid}@{\ }P{\colwidd}@{\ }P{\colwidd}@{\ }P{\colwidd}@{\ }P{\colwidd}}
\toprule
&&\multicolumn{7}{c}{Experimental configurations}\\
&&{\bf A}&{\bf B}&{\bf C}&{\bf D}&{\bf E}&{\bf F}&{\bf F}$_C$\\\cmidrule(l){2-9}
\multirow{2}{*}{Partitioning}&&\multirow{2}{*}{DCT}&\multirow{2}{*}{DCT}&\multirow{2}{*}{DCT}&VVC&VVC&VVC&VVC\\
&&&&&(square only)&(square only)&(square only)&(square only)\\\cmidrule(l){2-9}
\multirow{4}{2.5cm}{Transform core on $N{\times}N$ blocks} & \multirow{2}{*}{$N=8$} &\multirow{2}{*}{DCT}&\multirow{2}{*}{SBGFTs}&\multirow{2}{*}{SBGFTs}&\multirow{2}{*}{VVC}&\multirow{2}{*}{SBGFTs}&\multirow{2}{*}{SBGFTs}&SBGFTs\\
&&&&&&&&(top $C$)\\[4pt]
 & \multirow{2}{*}{$N>8$} &\multirow{2}{*}{DCT}&\multirow{2}{*}{DCT}&\multirow{2}{*}{SBGFTs}&\multirow{2}{*}{VVC}&\multirow{2}{*}{VVC}&\multirow{2}{*}{SBGFTs}&SBGFTs\\
 &&&&&&&&(top $C$)\\
 \bottomrule
 \end{tabular}
 \label{tab:conf}
\end{table*}

After selecting the optimal GFT $\mathcal{G}_{i}$ through \eqref{eq:lagrangian}, the quantized coefficients $\mathbf{\hat{X}}_j$ undergo arithmetic coding and are transmitted to the decoder, along with the corresponding transform index $j$ to ensure the correct inverse transformation can be applied at the decoder.

\subsection{Experiments on pixel domain}
\label{subsec:firstrun}
\noindent
In this section, we aim to provide an insight into the approximation capabilities of the SBGFTs family in the pixel domain. We evaluate the performance on a pair of commonly used datasets, the Miscellaneous set of the USC-SIPI Image Database \cite{usc-sipi} and the Kodak database \cite{kodak}, which include $22$ uncompressed images of size $512{\times}512$ and $24$ images of size $768{\times}512$, respectively.

To help navigate through the experiments, the examined configurations are named using letters from {\bf A} to {\bf F}, which are summarized in Table \ref{tab:conf} for easy referral.
Three different experimental configurations with the same partitioning strategy are examined first: the baseline {\bf A} (DCT only), {\bf B} (DCT and SBGFTs for $8 {\times} 8$ blocks only), and {\bf C} (variable size SBGFTs). In particular, in configuration {\bf B} we employ the SBGFTs when processing $8{\times}8$ blocks, while leaving the DCT to process $16 \times 16$ and $32 \times 32$ blocks. 
This configuration is especially relevant since its in-between structure allows us to assess the gain in compression performance when introducing variable size SBGFTs, supplementing $8{\times}8$-only SBGFTs that were previously considered in \cite{Gnutti_icip19, gnutti21, gnutti2024symmetric}.

\begin{figure}[t]
\centering
	\subfloat[USC dataset.]{\includegraphics[width=0.7\columnwidth]{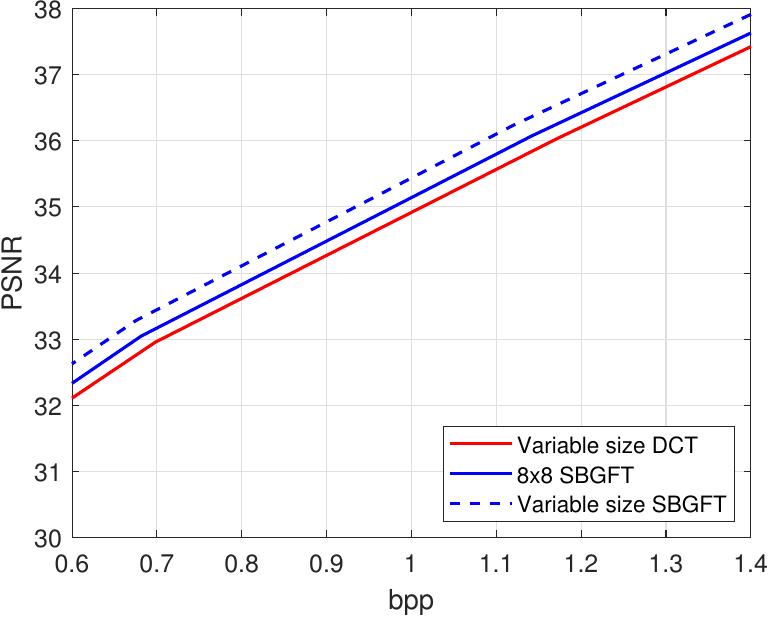}}\\
    \subfloat[Kodak dataset.]{\includegraphics[width=0.7\columnwidth]{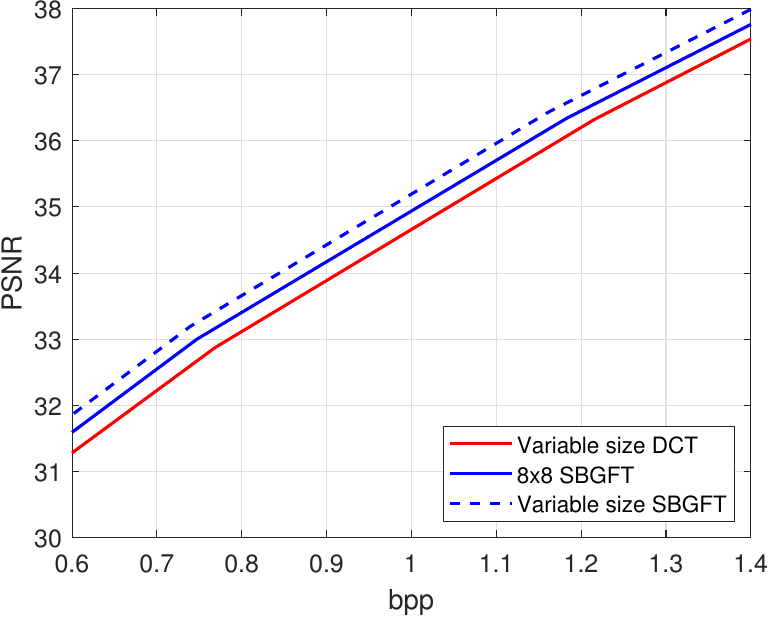}}
\caption{Performance comparison between the DCT and the SBGFTs set based transform cores in the pixel domain. Red: baseline, DCT-based configuration {\bf A}; Solid and dashed blue: configurations {\bf B} and {\bf C}, respectively.}
\label{fig:sbgftVSdct}
\end{figure}

The DCT is used to choose the optimal block partitioning for these three configurations.
%
Fixing a desired QP, the selected partition is the one that minimizes the following RD cost:
\begin{equation}
\label{eq:RD-partition}
\underset{\text{p}\in\mathcal{P}}{\min}\ J_\text{p} = D_\text{p}+\lambda \cdot R_\text{p}\,,
\end{equation}
in which the distortion and the rate are computed for each possible configuration (p) in the set of possible quad-tree partitioning $\mathcal{P}$.

Once the DCT-driven optimal partition of the macro-blocks is established, we then employ for each block size the transform sets prescribed by each configuration.
%
Note that this may not embody the optimal partition choice for the SBGFTs used in configuration {\bf B} and {\bf C}. However, this partitioning approach is much more practical because, since we are dealing with several sets of multiple transforms, the quad-tree exhaustive search would be of prohibitive complexity. Despite the partitioning being advantageous to configuration {\bf A}, the results reported in what follows indicate that SBGFTs still win out.
\begin{table}[t]
\caption{BD rate for Fig.~\ref{fig:sbgftVSdct}, configuration {\bf A} is the baseline.}
\centering
\begin{tabular}{lrrrr}
\cmidrule[.8pt]{2-5}
               & \multicolumn{2}{c}{USC dataset}            & \multicolumn{2}{c}{Kodak dataset}          \\ \cmidrule(r){2-5}
               Configuration & \multicolumn{1}{c}{{\bf B}\phantom{\%}} & \multicolumn{1}{c}{{\bf C}\phantom{\%}} & \multicolumn{1}{c}{{\bf B}\phantom{\%}} & \multicolumn{1}{c}{{\bf C}\phantom{\%}} \\ \midrule
\rowcolor[HTML]{C0C0C0} 
 $\Delta$ PSNR &  0.20\phantom{\%}               & 0.49\phantom{\%}                 &  0.27\phantom{\%}                & 0.53\phantom{\%}                \\
 $\Delta$ rate &  -3.33\%             &  -8.38\%            &   -3.94\%           & -7.76\%              \\ \bottomrule
 \end{tabular}
 \label{tab:Bjo}
\end{table}
\begin{figure*}[p]
    \centering
        \subfloat[configuration {\bf B} ($\mbox{QP}=25$).]{\includegraphics[width=0.24\textwidth]{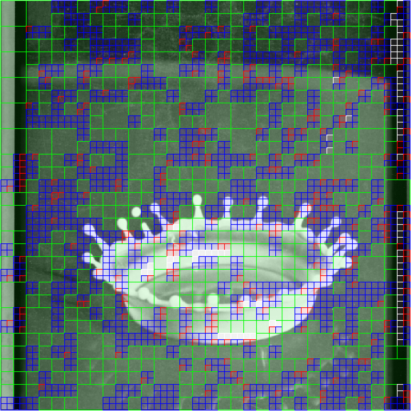}} \
        \subfloat[configuration {\bf B} ($\mbox{QP}=30$).]{\includegraphics[width=0.24\textwidth]{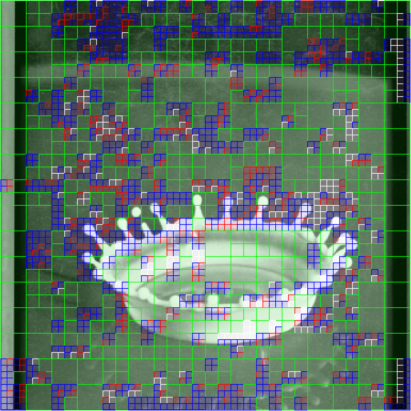}} \
        \subfloat[configuration {\bf B} ($\mbox{QP}=35$).]{\includegraphics[width=0.24\textwidth]{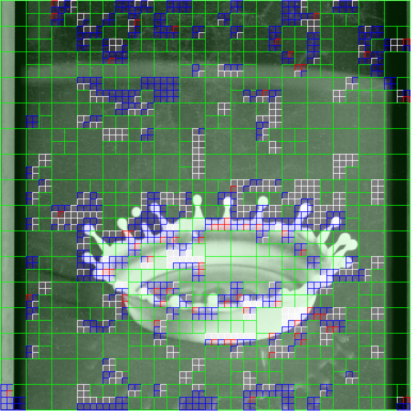}} \
        \subfloat[configuration {\bf B} ($\mbox{QP}=40$).]{\includegraphics[width=0.24\textwidth]{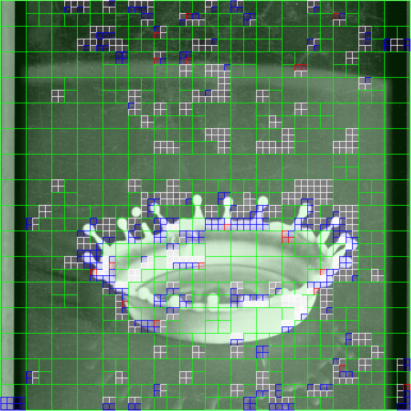}}\\
        \subfloat[configuration {\bf C} ($\mbox{QP}=25$).]{\includegraphics[width=0.24\textwidth]{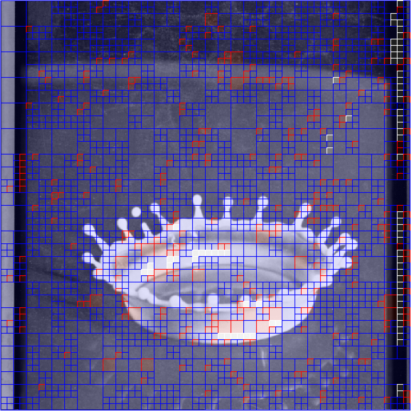}} \
        \subfloat[configuration {\bf C} ($\mbox{QP}=30$).]{\includegraphics[width=0.24\textwidth]{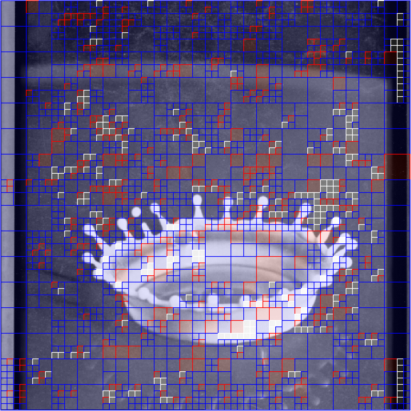}} \
        \subfloat[configuration {\bf C} ($\mbox{QP}=35$).]{\includegraphics[width=0.24\textwidth]{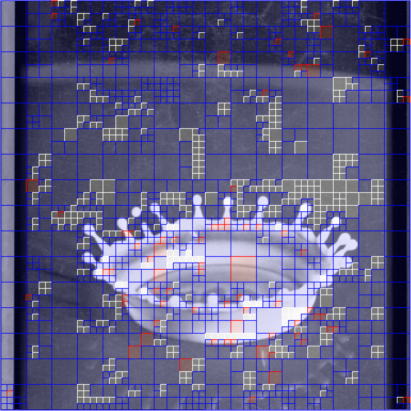}} \
        \subfloat[configuration {\bf C} ($\mbox{QP}=40$).]{\includegraphics[width=0.24\textwidth]{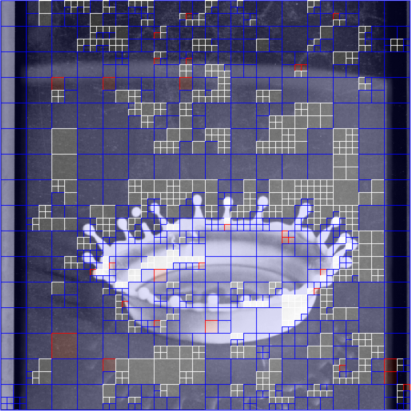}}\\
        \subfloat[configuration {\bf B} ($\mbox{QP}=25$).]{\includegraphics[width=0.24\textwidth]{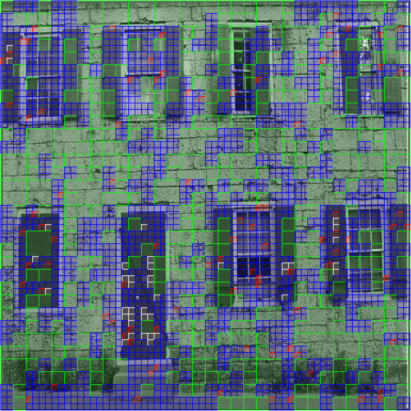}} \
        \subfloat[configuration {\bf B} ($\mbox{QP}=30$).]{\includegraphics[width=0.24\textwidth]{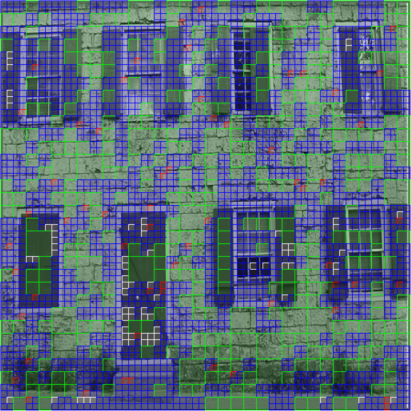}} \
        \subfloat[configuration {\bf B} ($\mbox{QP}=35$).]{\includegraphics[width=0.24\textwidth]{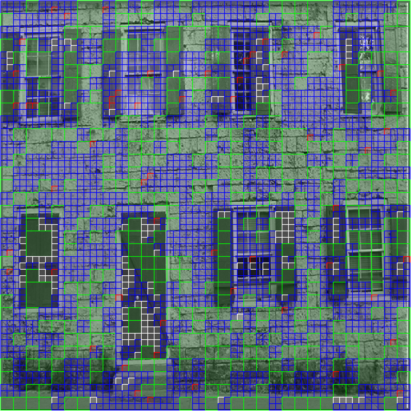}} \
        \subfloat[configuration {\bf B} ($\mbox{QP}=40$).]{\includegraphics[width=0.24\textwidth]{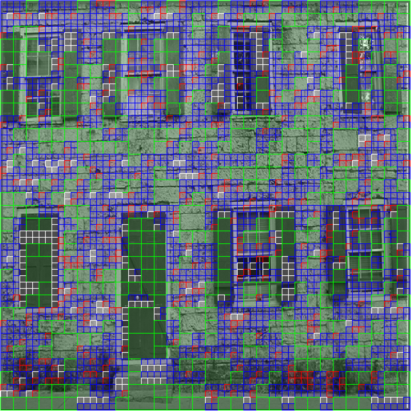}}\\
        \subfloat[configuration {\bf C} ($\mbox{QP}=25$).]{\includegraphics[width=0.24\textwidth]{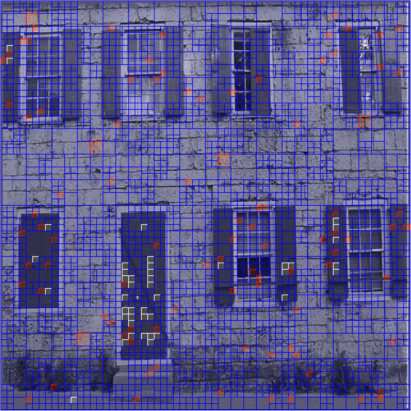}} \
        \subfloat[configuration {\bf C} ($\mbox{QP}=30$).]{\includegraphics[width=0.24\textwidth]{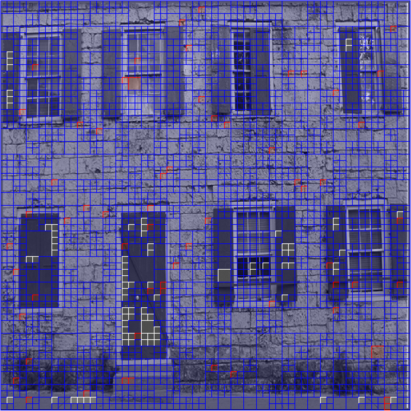}} \
        \subfloat[configuration {\bf C} ($\mbox{QP}=35$).]{\includegraphics[width=0.24\textwidth]{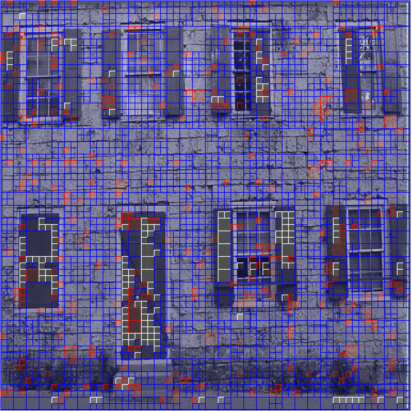}} \
        \subfloat[configuration {\bf C} ($\mbox{QP}=40$).]{\includegraphics[width=0.24\textwidth]{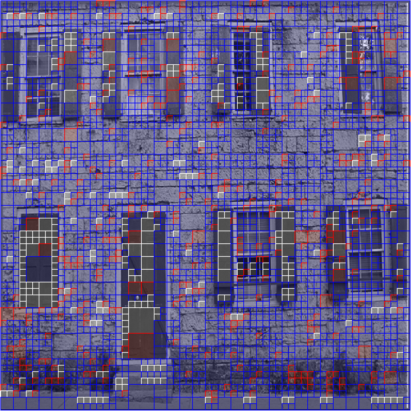}}
    \caption{Visual illustration of the best performing transforms for each block based on the RDOT paradigm in the pixel domain. Recall that block partitioning is dependent on the QP. The first and third rows correspond to configuration {\bf B} where only the $8{\times}8$ SBGFTs are employed in place of the $8{\times}8$ DCT. Blue and red indicate whether the best transform is one of the SBGFTs as per configuration {\bf B} or the DCT as per configuration {\bf A}, respectively. Green identifies larger blocks, where the DCT is applied by default. In the second and fourth rows, the configuration {\bf C} is used, where the SBGFTs are employed for every block size. In this case there are no green blocks, as the best transform among the optimal SBGFT or the baseline DCT for every block size is always assessed. For blocks colored in white, only the DC coefficient survives after quantization for each transform, thus there is no difference in whichever transform is employed, which is a more common occurrence for low rates (high QP).}
    \label{fig:visualComparison}
\end{figure*}

Arithmetic coding is applied to generate the bitstream, encoding both the quantized transform coefficients and, for the SBGFTs, the graph index $j$ as side information to specify the selected graph. In these experiments, we employ a simple fixed-length code of $\lceil \log_2(l) \rceil$ bits to represent the graph index used for a given $N{\times}N$ block, with $l=8N-24$.

Fig.~\ref{fig:sbgftVSdct} illustrates the RD performance of
configurations A, B, and C for both still image datasets, plotting PSNR versus bpp in terms of bitstream length generated by arithmetic encoding. 
In our experiments, we consider QP values of $25$, $30$, $35$, $40$, and $45$. Fig.~\ref{fig:sbgftVSdct} shows that replacing the $8{\times}8$ DCT with $8{\times}8$ SBGFTs already results in a sizable performance gain, which is further increased when also implementing the SBGFTs sets for blocks of different size within the compression framework.


In Table \ref{tab:Bjo}, the corresponding Bj\o ntegaard's metric is also reported to show the average gain in terms of PSNR and the average bit rate saving percentage between the three examined transform configurations.
On both datasets, configuration {\bf B} already has a sizable coding gain with respect to the DCT, and configuration {\bf C} approximately double that gain. 

To further showcase the SBGFTs coding efficiency, a visually illustrative comparison is presented in Fig.~\ref{fig:visualComparison}, which displays the best transform based on the RDOT for each block in the partition obtained for various QPs, thus at different rates. The comparison is done for two reference images, one for each dataset.
%
%
These results indicate that, regardless of the quantization level, the SBGFTs outnumber the DCT for both configurations {\bf B} and {\bf C}.


\subsection{Experiments on residuals}
\label{subsec:secondrun}
\noindent
The next set of experiments is performed on residual data derived from video intra-prediction using the All-Intra (AI) profile of VVC. The objective is to evaluate how SBGFTs perform on residuals against the DCT variants used within the VVC framework, using square blocks only. 
In particular, we have extracted a total of $483 \, 067$, variable size, square residual blocks from $4$ standard test video signals: \textit{BQMall}, \textit{BasketballDrill}, \textit{Mobcal}, and \textit{Shields}.

Two experimental configurations {\bf D} and {\bf E} using (at least partially) the VVC primary transforms are to be compared with configuration {\bf F} using only variable size SBGFTs, all employing VVC partitioning. To ensure that fair comparisons are made, when considering {\bf D} and {\bf E}, the VVC Explicit Multiple Transform Selection (Explicit MTS) method is employed. Explicit MTS is the most complex encoding profile since it is an exhaustive RDOT scheme that includes all four additional primary transforms (DCT-VIII, DST-VII, and their horizontal and vertical combinations) in addition to the standard type-$2$ DCT. Therefore, Explicit MTS is akin to the scheme depicted in Fig.~\ref{fig:overview_standard} with $l=5$ transforms. The secondary, non-separable VVC transform, which further adds to encoding complexity, is not relevant to this study since it is designed to compact VVC primary transform coefficients further, and thus, it is not applicable to SBGFTs. The previously considered QPs have been employed again for these experiments.

\begin{figure}[t]
\centering
	\includegraphics[width=0.7\columnwidth]{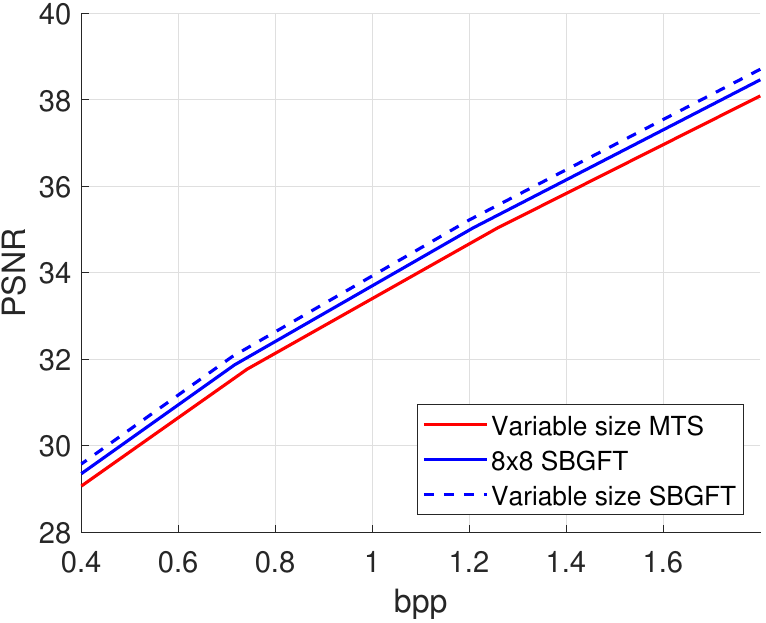}
    \caption{Performance comparison between the VVC Explicit MTS method and the SBGFTs sets in the residuals domain. Red: configuration {\bf D} (variable size, Explicit MTS primary transforms); Solid and dashed blue: configuration {\bf E} ($8{\times}8$ SBGFTs only) and {\bf F} (variable-size SBGFTs), respectively.}
\label{fig:sbgftVSMTS}
\end{figure}
\begin{table}[t]
\centering
\captionof{table}{BD rate associated with Fig.~\ref{fig:sbgftVSMTS}, and percentage increase in time complexity for encoding ($\Delta T_{\mathcal{E}}$) and decoding ($\Delta T_{\mathcal{D}}$), considering configuration {\bf D} as baseline, when applying configurations {\bf E}, {\bf F} and {\bf F}$_5$.}
\begin{tabular}{lrrr}

 \toprule
               Configuration & \multicolumn{1}{c}{{\bf E}\phantom{\%}} & \multicolumn{1}{c}{{\bf F\phantom{\%}}} & \multicolumn{1}{c}{{\bf F}$_5$\phantom{\%}} \\ \midrule
 \rowcolor[HTML]{C0C0C0} 
 $\Delta$ PSNR & 0.28\phantom{\%}                & 0.52\phantom{\%}                 & 0.38\phantom{\%}                                 \\
 $\Delta$ rate &  -5.14\%             & -9.29\%             & -6.23\%                           \\\midrule
 $\Delta T_{\mathcal{E}}$& 2.5\% & 100.2\% & $\approx$ 0.0\%\\
 $\Delta T_{\mathcal{D}}$& $\approx$ 0.0\% & $\approx$ 0.0\% & $\approx$ 0.0\%\\\bottomrule
 \end{tabular}
 \label{tab:Bjo2}
 \end{table}

In baseline configuration {\bf D}, transform coding of any sized residual block is done using MTS transforms only, using the best among the five transforms in an RD sense. Such best transform is selected for each block by searching for the smallest Lagrangian cost, using the function in \eqref{eq:lagrangian} with:
\begin{equation}\label{eq:lambda}
\lambda = 0.57 \cdot 2^{(QP-12)/3}
\end{equation}
which mirrors the actual VVC rate-control function. The index of the optimal transform for each block is represented with the corresponding MTS flag adopted in the VVC standard.

Similarly to what was done in Sec.~\ref{subsec:firstrun}, to assess the gain in compression performance when using SBGFTs larger than $8{\times}8$, we introduce configuration {\bf E}. As in configuration {\bf B}, we only transform $8{\times}8$ blocks using the $8{\times}8$ SBGFTs set, while using VVC primary transforms for blocks of other sizes, where the Explicit MTS method is employed.

Fig.~\ref{fig:sbgftVSMTS} depicts the performance of these three transform configurations. It clearly shows that substituting the $8{\times}8$ DCT variants with $8{\times}8$ SBGFTs already achieve better performance, which is further improved when integrating variable size SBGFTs within the framework. The performance in terms of BD rate is shown in Table \ref{tab:Bjo2}, which is further discussed in Sec.~\ref{subsec:low_complexity}.

\begin{figure*}[p]
    \centering
        \subfloat[$16{\times}16$ SBGFT (PM=0).]{\includegraphics[width=0.21\textwidth]{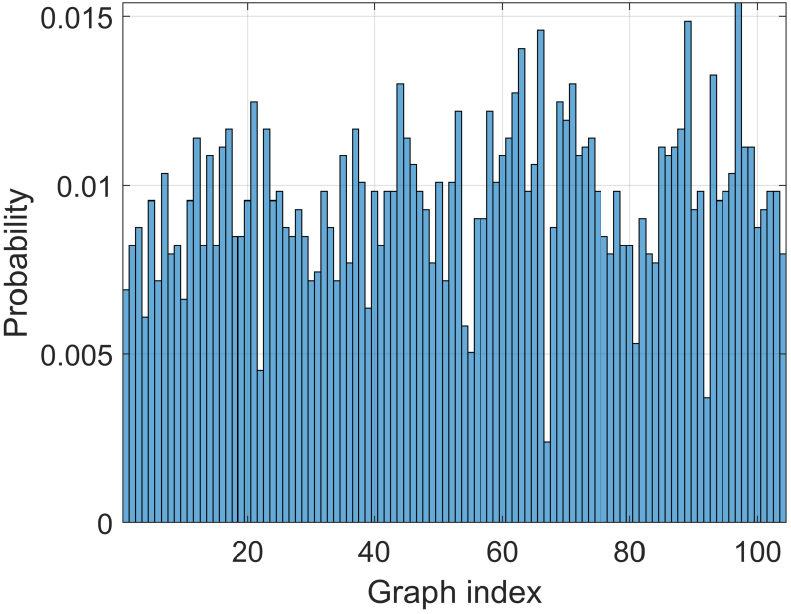}} \qquad
        \subfloat[$16{\times}16$ SBGFT (PM=1).]{\includegraphics[width=0.21\textwidth]{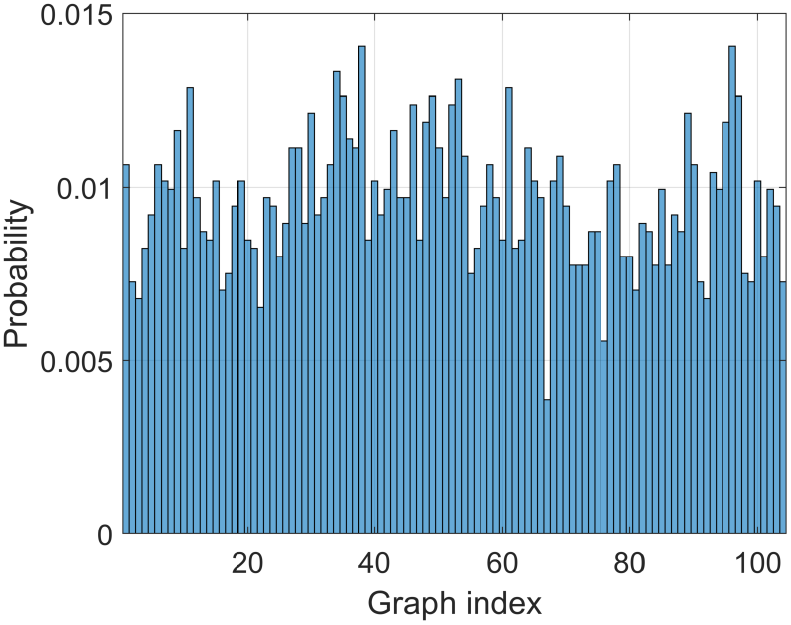}} \qquad
        \subfloat[$16{\times}16$ SBGFT (PM=2).]{\includegraphics[width=0.21\textwidth]{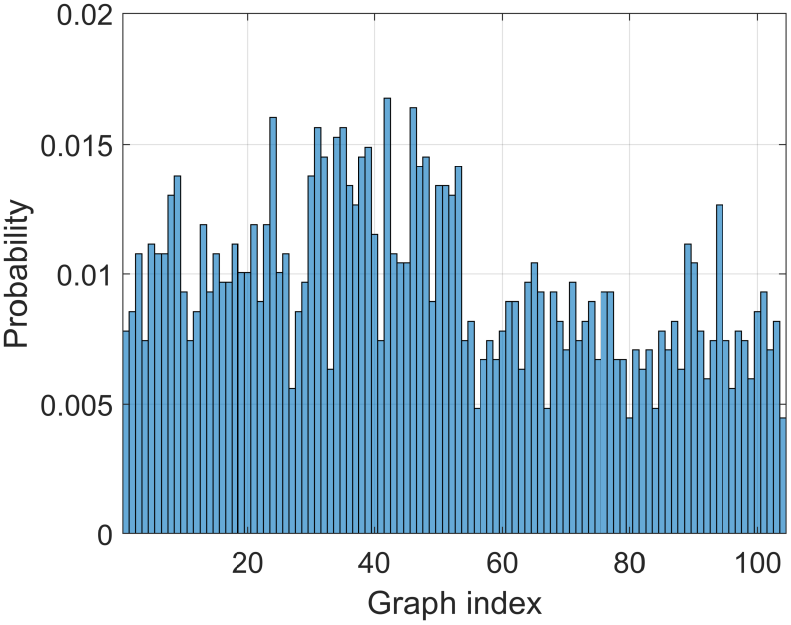}} \qquad
        \subfloat[$16{\times}16$ SBGFT (PM=3).]{\includegraphics[width=0.21\textwidth]{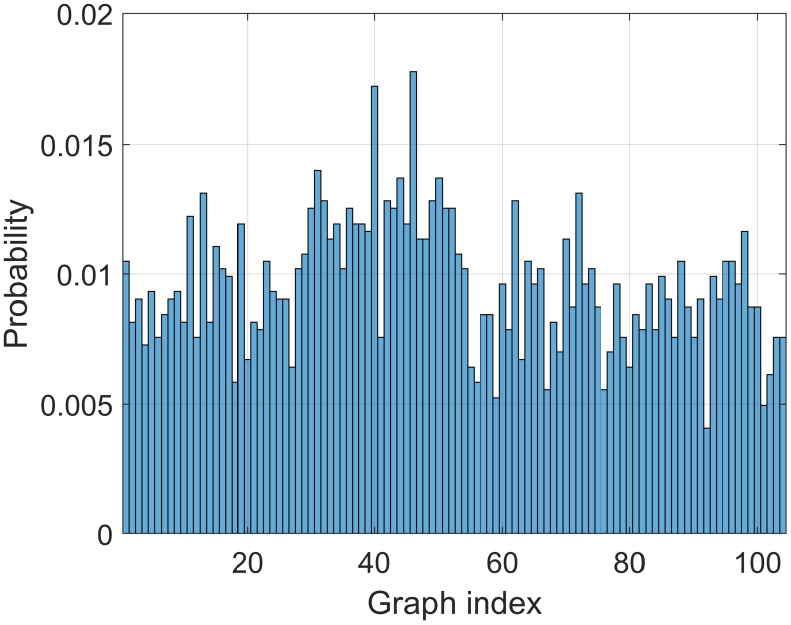}} \\
        \subfloat[$16{\times }16$ SBGFT (PM=4).]{\includegraphics[width=0.21\textwidth]{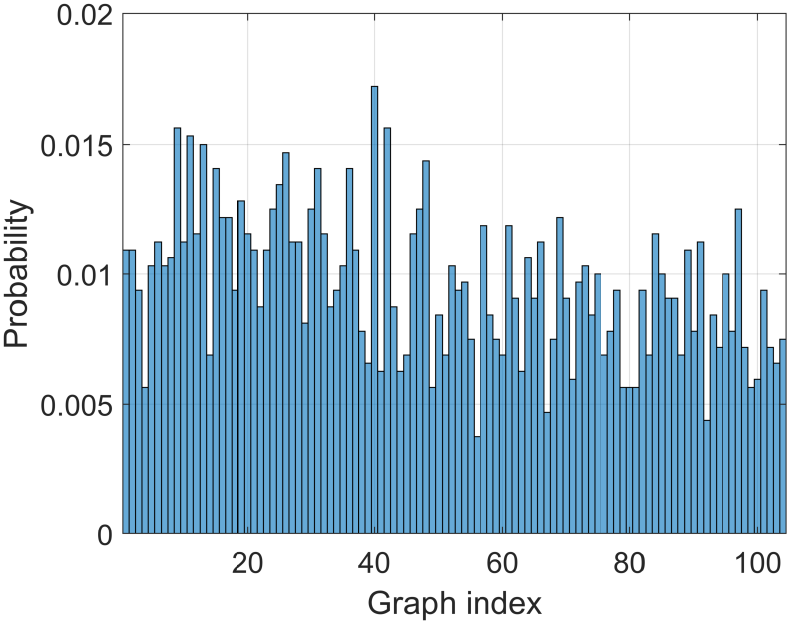}} \qquad
        \subfloat[$16{\times }16$ SBGFT (PM=5).]{\includegraphics[width=0.21\textwidth]{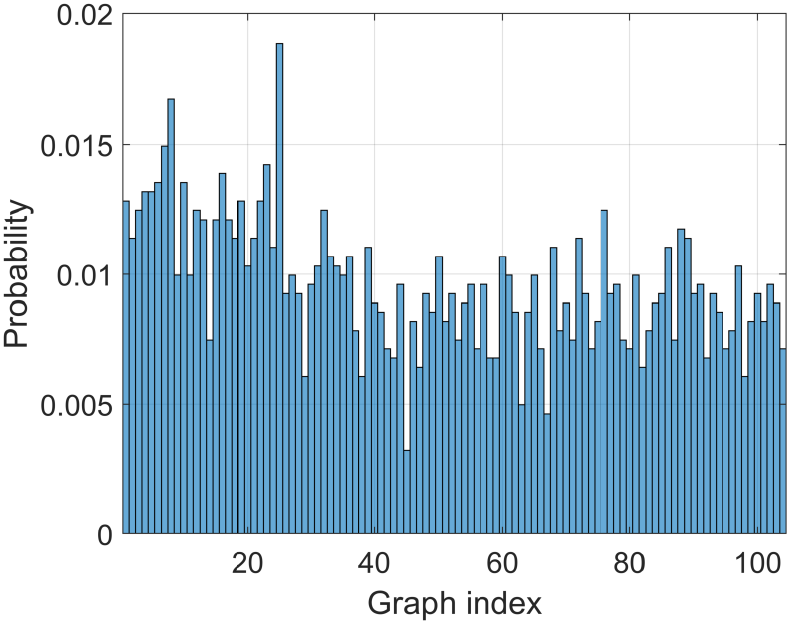}} \qquad
        \subfloat[$16{\times }16$ SBGFT (PM=6).]{\includegraphics[width=0.21\textwidth]{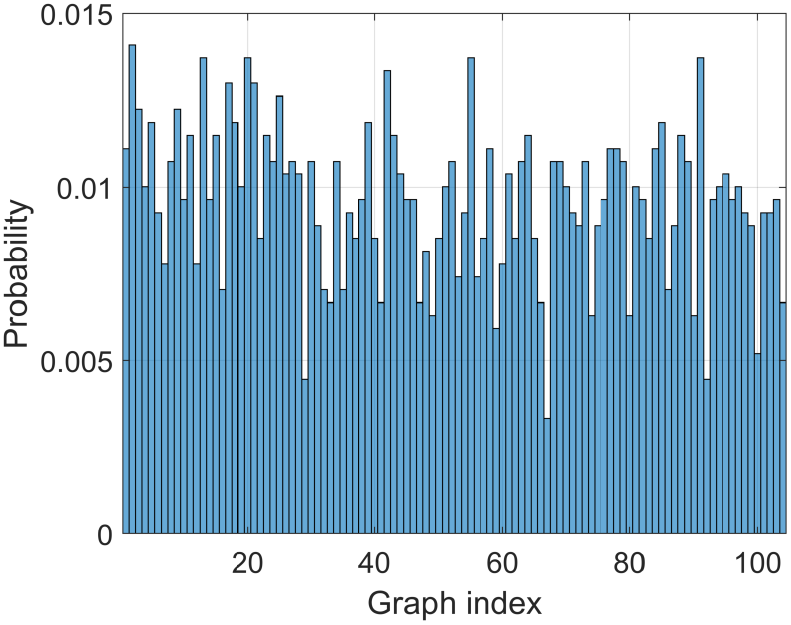}} \qquad
        \subfloat[$16{\times}16$ SBGFT (PM=7).]{\includegraphics[width=0.21\textwidth]{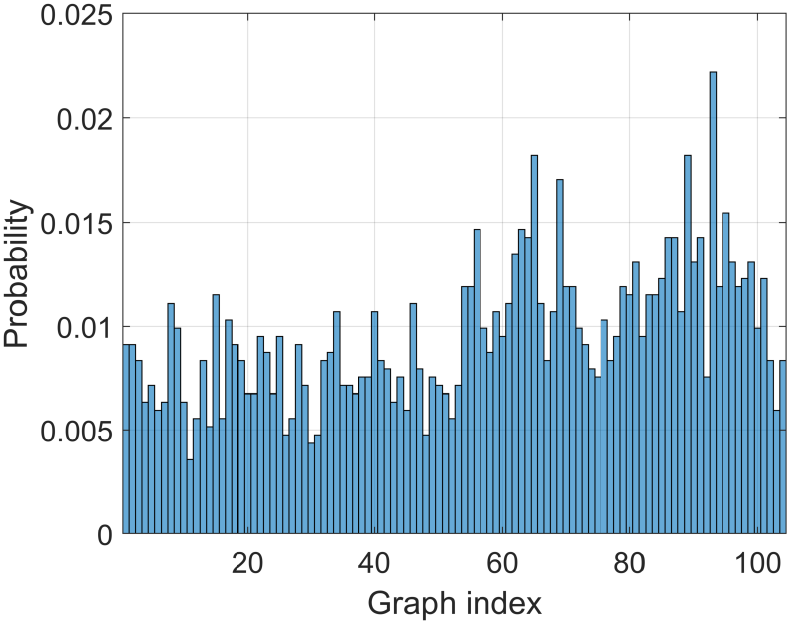}} \\
        \subfloat[$32{\times}32$ SBGFT (PM=0).]{\includegraphics[width=0.21\textwidth]{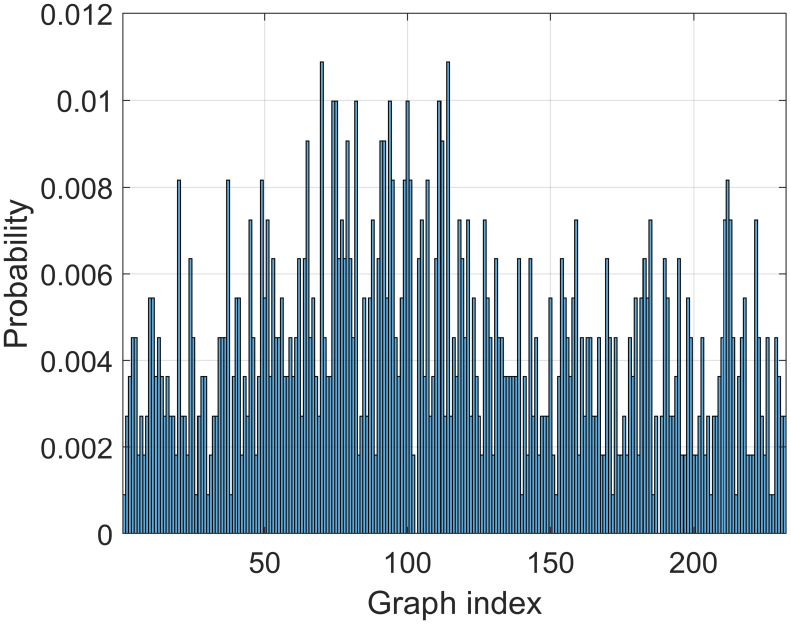}} \qquad
        \subfloat[$32{\times}32$ SBGFT (PM=1).]{\includegraphics[width=0.21\textwidth]{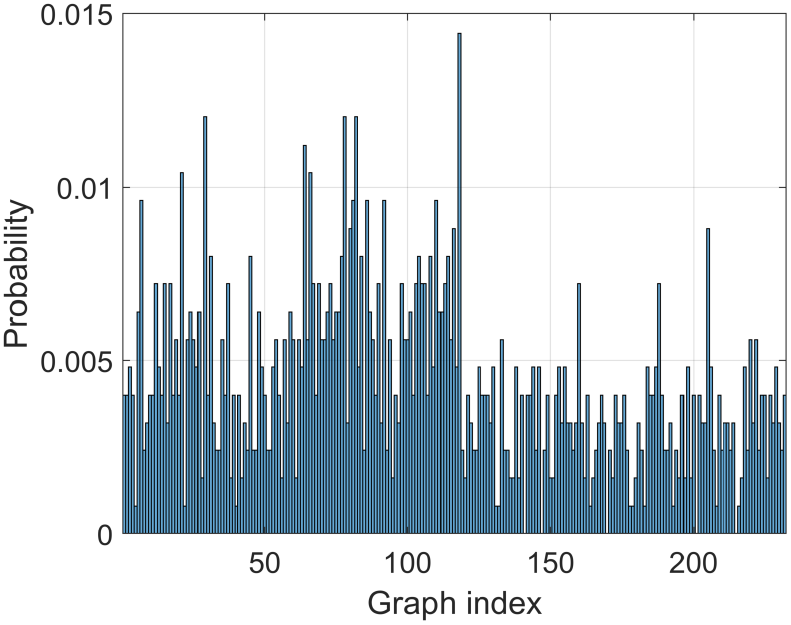}} \qquad
        \subfloat[$32{\times}32$ SBGFT (PM=2).]{\includegraphics[width=0.21\textwidth]{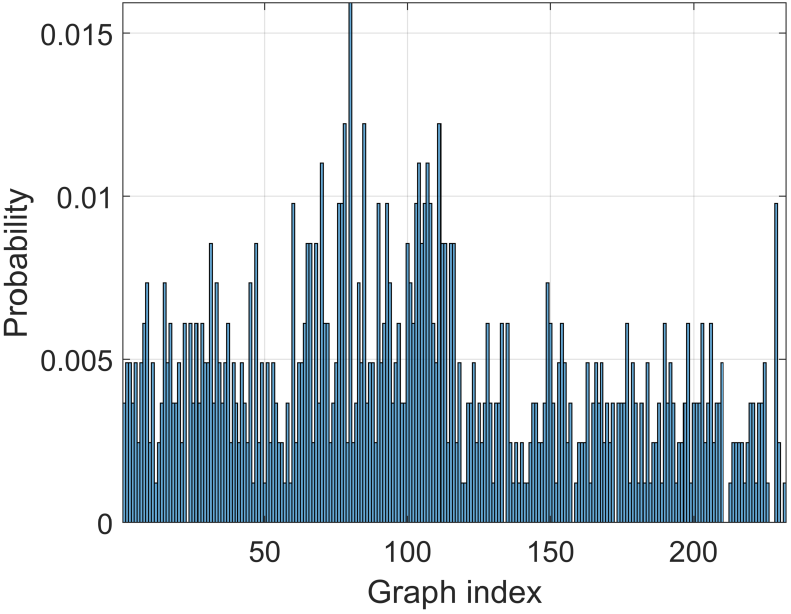}} \qquad
        \subfloat[$32{\times}32$ SBGFT (PM=3).]{\includegraphics[width=0.21\textwidth]{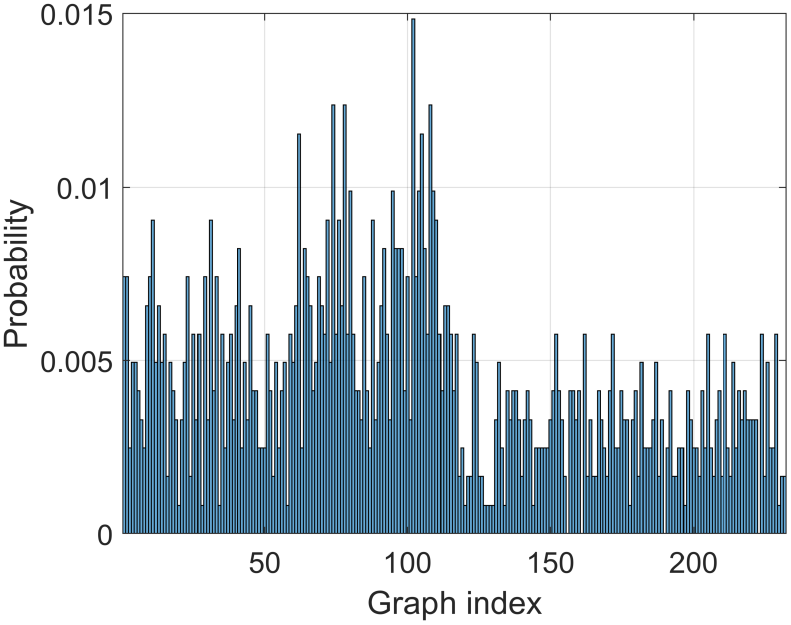}} \\
        \subfloat[$32{\times}32$ SBGFT (PM=4).]{\includegraphics[width=0.21\textwidth]{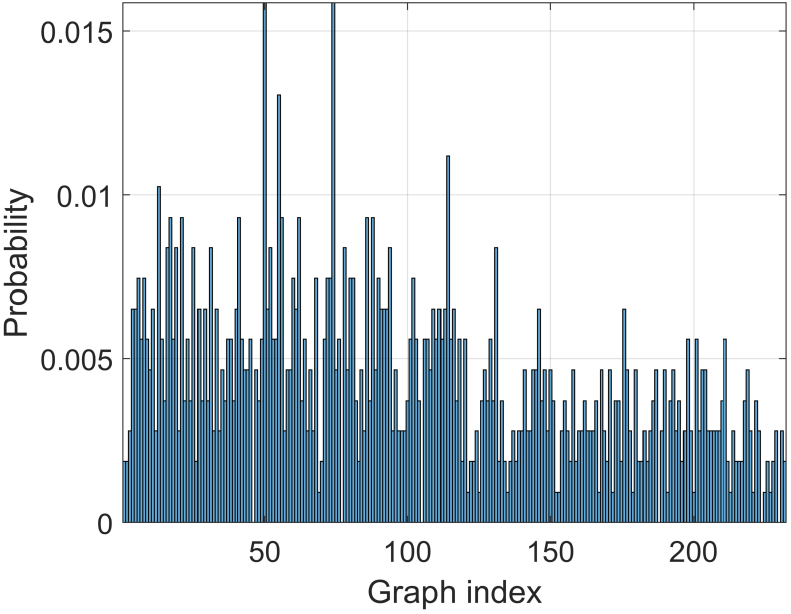}} \qquad
        \subfloat[$32{\times}32$ SBGFT (PM=5).]{\includegraphics[width=0.21\textwidth]{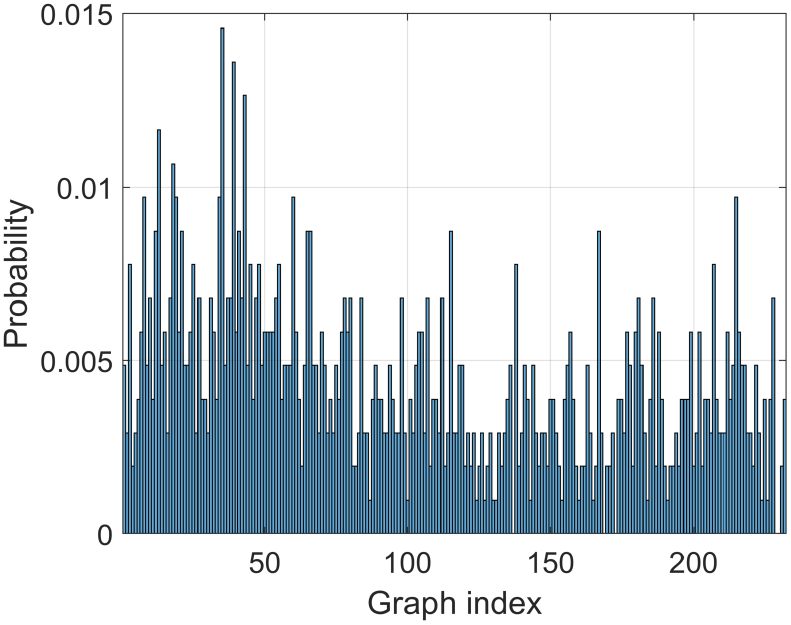}} \qquad
        \subfloat[$32{\times}32$ SBGFT (PM=6).]{\includegraphics[width=0.21\textwidth]{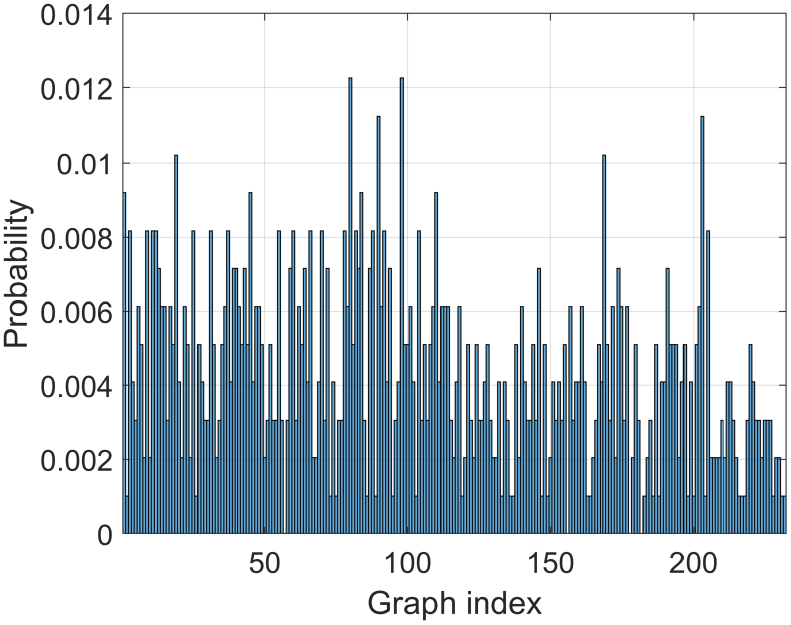}} \qquad
        \subfloat[$32{\times}32$ SBGFT (PM=7).]{\includegraphics[width=0.21\textwidth]{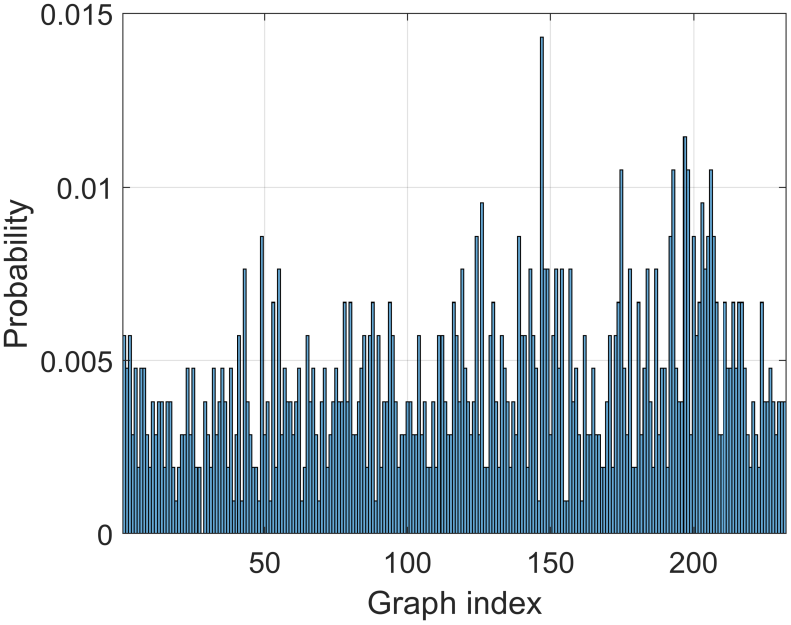}} \\       
    \caption{Probability distributions of optimal graphs for a subset of prediction modes. The first two rows correspond to $16{\times}16$ blocks, while the last two are associated with $32{\times}32$ blocks. The depicted distributions are related to QP $= 25$.\label{fig:stats}}
    \subfloat[Low complexity variable size SBGFTs performance]{\includegraphics[width=0.7\columnwidth]{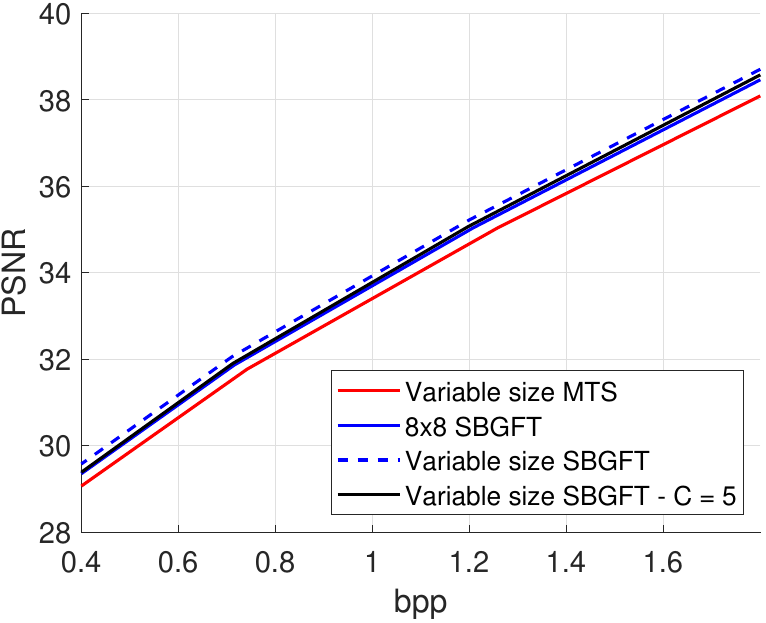}}\qquad\qquad 
    \subfloat[Zoom of (a)]{\includegraphics[width=0.7\columnwidth]{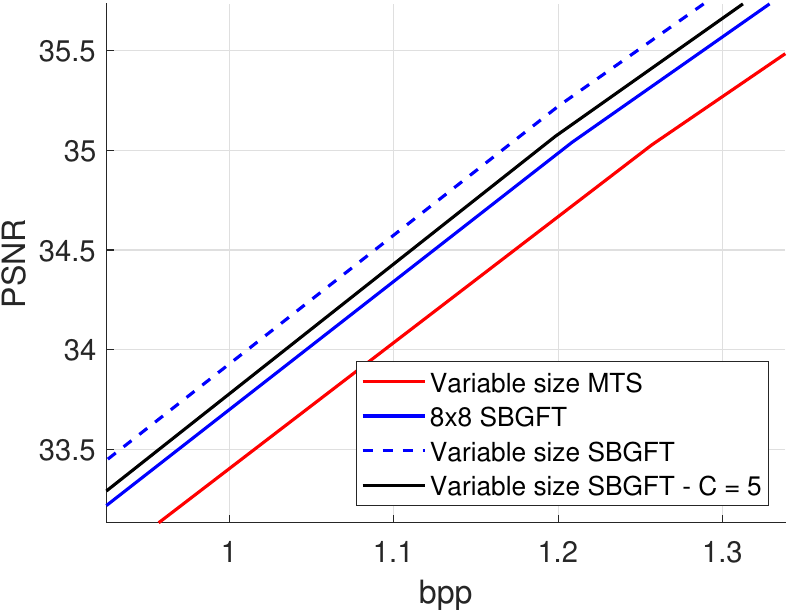}}
    \caption{Performance comparison between configuration {\bf F}$_5$ and the other configurations on residuals. Red: configuration {\bf D} (variable size, Explicit MTS primary transforms); Solid and dashed blue: configuration {\bf E} ($8{\times}8$ SBGFTs only) and {\bf F} (variable-size SBGFTs), respectively. Black: configuration {\bf F}$_5$ (variable-size SBGFTs using only the top $5$ transforms).\label{fig:res_subset}}
\end{figure*}


\subsection{Complexity analysis}
\label{subsec:complexity_analysis}
\noindent
In this section, we provide an analysis of the computational complexity linked to using variable size SBGFTs as the transform core in the experiments carried out on residuals.
Table \ref{tab:Bjo2} shows the time complexity increase of configurations {\bf E} and {\bf F} relative to the baseline configuration {\bf D} (Explicit MTS) in terms of relative decoding and encoding time increase $\Delta T_{\mathcal{D}}$ and $\Delta T_{\mathcal{E}}$, respectively.

First, we can observe that no decoding time penalty is expected. In the examined configurations, the only difference in the block reconstruction stage is to employ an inverse SBGFT instead of the inverse of a VVC primary transform, all of which admit fast implementations. Thus, there is no noticeable increase $\Delta T_{\mathcal{D}}$ in the decoding time (see Table \ref{tab:Bjo2}).

For the encoding stage, $\Delta T_{\mathcal{E}}$ is calculated by considering the average time spent on the entire encoding process, including both block coding transforms and the other computationally intensive tasks, such as intra-prediction and block partitioning.

In particular, the encoding time on the average increases by a slight $\Delta T_{\mathcal{E}}=2.5\%$ margin in configuration {\bf E} with respect to configuration {\bf D}, which is due to the exhaustive search on $40$ fast transforms instead of just $5$ as in VVC Explicit MTS for $8{\times}8$ blocks. The result is within the expected range when considering that the SBGFTs are actually applied on about $40\%$ of the frame area occupied by $8{\times}8$ blocks.

However, the encoding time for configuration {\bf F} is about twice that of configuration {\bf D}. In fact, the exhaustive RDOT search becomes much more demanding for variable size SBGFTs since there are $8N-24$ transforms to be computed for any given $N{\times}N$ block. Such high complexity confirms that an exhaustive search among the potential transforms is computationally inefficient, and it causes unsatisfactorily higher encoding computational times.

To reduce the encoding complexity of the proposed variable size SBGFTs framework, 
in the next section, we propose a low-complexity implementation that brings the encoding time back to that of configuration {\bf E} (first row of Table \ref{tab:Bjo2}) without sacrificing much in terms of coding performance.

\subsection{Low complexity variable size SBGFTs framework}
\label{subsec:low_complexity}
\noindent
In this section, we investigate the correlation between VVC Prediction Modes (PMs) and optimal SBGFTs, so that we can leverage these findings to design a low complexity variable size SBGFTs framework. This framework entails employing a predetermined restricted set of SBGFTs, instead of the whole set of $8N{-}24$ graphs. The SBGFTs of this subgroup are selected given the PM from which a given block is derived.

Hence, we begin by taking a secondary dataset of video frames including $3$ standard test video signals: \textit{PartyScene}, \textit{RaceHorsesC}, and \textit{Traffic}, which differ from those used in the previous experiments to exclude bias in the subsequent validation. We partition the visual data into variable-sized residual blocks as we did in Sec.~\ref{subsec:secondrun}. Subsequently, we apply all the SBGFTs for each variable-sized block and determine the optimal SBGFT with the standard RD optimization process.

Next, we perform the following analysis: we gather all residual blocks of the same size derived from the same prediction mode and then calculate the relative frequency with which a particular SBGFT is selected as optimal among the entire corresponding SBGFTs set. This process is repeated for all block sizes and all prediction modes, varying QP as usual.



Fig.~\ref{fig:stats} presents a selection of histograms giving the relative frequency of each SBGFT on the secondary dataset (the full histogram collection is available as supplementary material). These histograms reveal two main findings: (i) the relative frequency distributions of the various prediction modes are generally non-uniform, and (ii) these distributions vary across different PMs. These observations imply that there is indeed a correlation between prediction modes and optimal SBGFTs. In particular, (i) suggests that, for a given PM, we may be able to limit the number of graphs to consider within the multiple transforms framework. Furthermore, (ii) indicates that all SBGFTs within the set are useful, at least for some PMs.


Thus, we now modify our proposed SBGFTs-based configuration {\bf F} by introducing a low complexity experimental configuration {\bf F}$_C$ where, rather than evaluating all $8N {-} 24$ graphs for each $N {\times} N$ block, we only examine a restricted set of them with a designated cardinality $C_{\text{PM}}$, which in principle could be optimally designed for each PM. This means that a different subgroup of graphs that comprise the top $C_{\text{PM}}$ SBGFTs for each PM is used. The composition of this smaller set depends on the prediction mode from which a given block originates as it relies on the relative frequency distribution of the SBGFTs as inferred by the histogram. 


To exactly match the complexity of VVC for larger blocks as those found in configurations {\bf D} and {\bf E}, a reasonable approach is to choose $C_{\text{PM}}=C=5$ for every PM, so that the number of considered SBGFTs aligns with that of the primary transforms set employed in Explicit MTS. Therefore, we repeat the experiments performed in Sec.~\ref{subsec:secondrun} using configuration {\bf F}$_5$, where the RDOT paradigm is applied for all block sizes using only the top $5$ SBGFTs found for each PM.
Therefore, the complexity of configuration {\bf F}$_5$ matches that of configuration {\bf D} and is lower than that of configuration {\bf E} (and {\bf F}, naturally).

The RD performance for configuration {\bf F}$_5$ is plotted in Fig.~\ref{fig:res_subset}. Again, {\bf F}$_5$ is better than configuration {\bf D} (VVC Explicit MTS). It is also better than configuration {\bf E}, therefore showing again the better approximation capabilities of SBGFTs with respect to DCT variants for larger blocks, even if the optimal graph may not be part of the SBGFTs PM-derived subgroup. Since the latter event may indeed happen, there is a performance drop for configuration {\bf F}$_5$ with respect to configuration {\bf F}. These conclusions are valid across the entire considered rate spectrum.

Table \ref{tab:Bjo2} shows the corresponding BD rate to illustrate the average gain in terms of PSNR and the average bit rate saving percentage between the rate-distortion curves associated with the above-discussed configurations {\bf F} (full set, variable size SBGFTs), {\bf E} (full set, $8{\times}8$ SBGFTs plus Explicit MTS for $N>8$), and {\bf F}$_5$ (full set, restricted set with $C=5$ SBGFTs) with respect to configurations {\bf D} (Explicit MTS for all $N$). In the end, in our experiments, the low complexity configuration {\bf F}$_5$  achieves a significant bitrate saving of more than $6\%$ with respect to VVC Explicit Multiple Transform Selection mode.

\section{Conclusions}
\label{sec:concl}
\noindent
In this paper, we have expanded the notion of Symmetry-Based Graph Fourier Transforms (SBGFTs) from $8{\times}8$ grids to the case of general $N{\times}N$ grids. The SBGFTs are non-separable transforms with directional bases that provide sparse signal representation and high image compression performance while preserving low computational complexity thanks to their symmetry properties.

Then, we have proposed integrating variable-size SBGFTs into a coding framework, demonstrating that their RD performance surpasses both the traditional DCT-II and the Explicit Multiple Transform Selection (MTS) method of VVC. Additionally, we have explored correlations between optimal graphs and prediction modes to reduce the cardinality of the transform sets, thus presenting a low-complexity framework.

Even in the low complexity scenario, the proposed SBGFTs compression framework achieves a bit rate saving of $6.23\%$, with only a marginal increase in average encoding complexity. This suggests that the use of SBGFTs in an RDOT scheme has the potential to improve image and video compression.

\newpage
\bibliographystyle{IEEEtran}
\bibliography{refs}

\begin{IEEEbiography}
[{\includegraphics[width=1in,height=1.25in,clip,keepaspectratio]{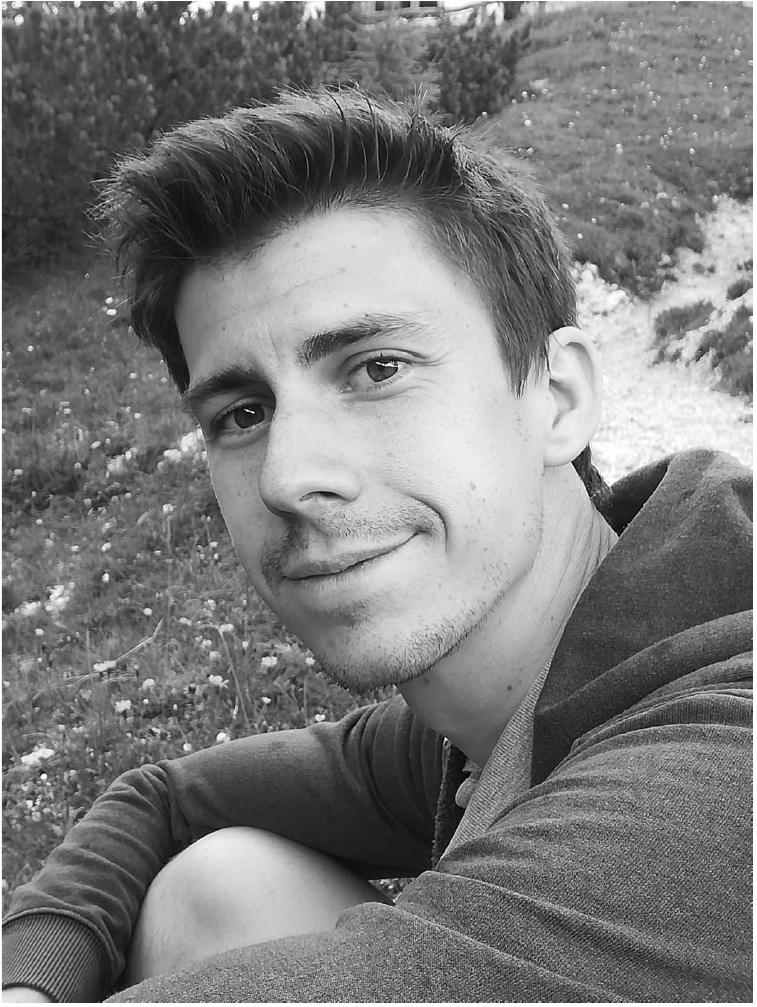}}]{Alessandro Gnutti} received the MS~degree (cum laude) in Telecommunications engineering and the Ph.D.~degree in Information engineering from the University of Brescia, Italy, in 2014 and 2017, respectively. He is currently tenure-track assistant professor with the Department of Information Engineering, University of Brescia. In 2017, he was a visiting fellow at the University of Southern California, working on graph signal processing. His main research interests cover signal and image representation, including transform and graph analysis for image and video compression, and pattern recognition.
\end{IEEEbiography}

\begin{IEEEbiography}
[{\includegraphics[width=1in,height=1.25in,clip,keepaspectratio]{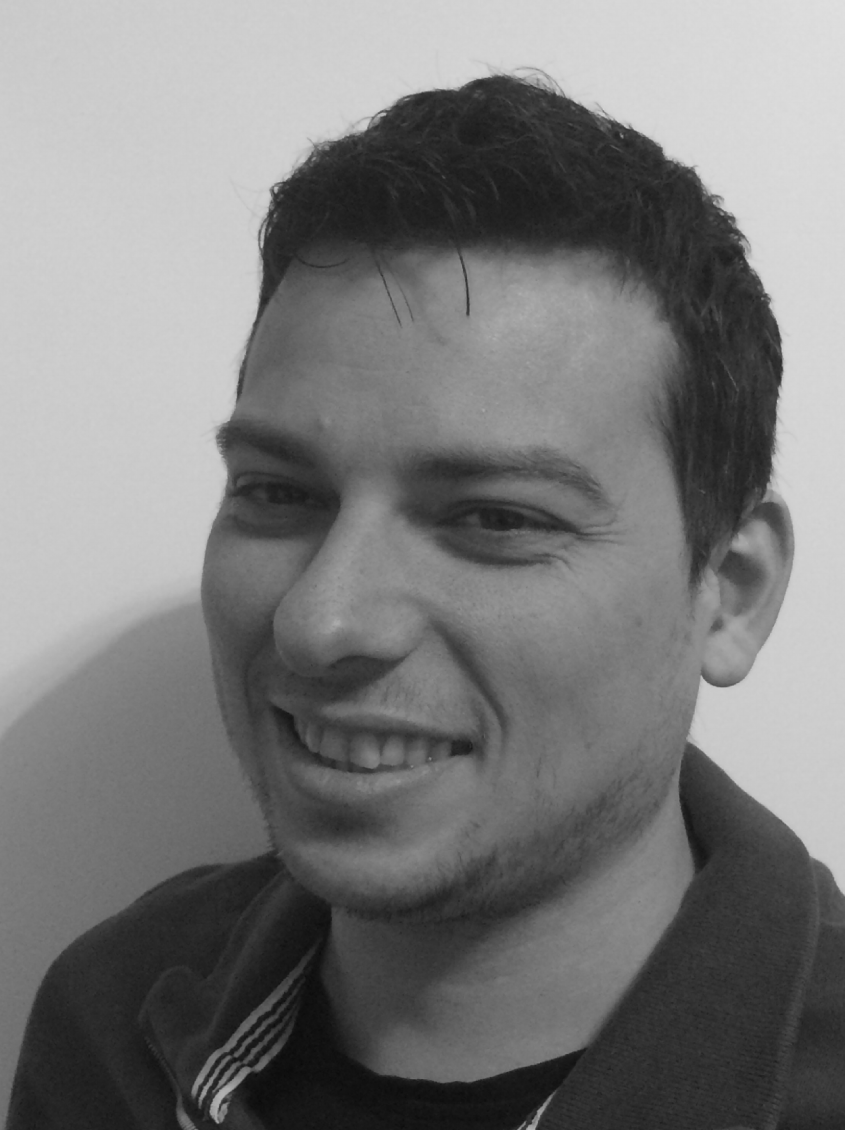}}]{Fabrizio Guerrini}
received the MS~degree (cum laude) in Electronic engineering and the Ph.D.~degree in Information engineering from the University of Brescia, Italy, in 2004 and 2008, respectively. He is currently a tenure-track assistant professor with the Department of Information Engineering, University of Brescia. His main research interests cover image and video processing and applications, including compression, transform coding and feature extraction, image security and watermarking, and pattern analysis and recognition.
\end{IEEEbiography}

\begin{IEEEbiography}
[{\includegraphics[width=1in,height=1.25in,clip,keepaspectratio]{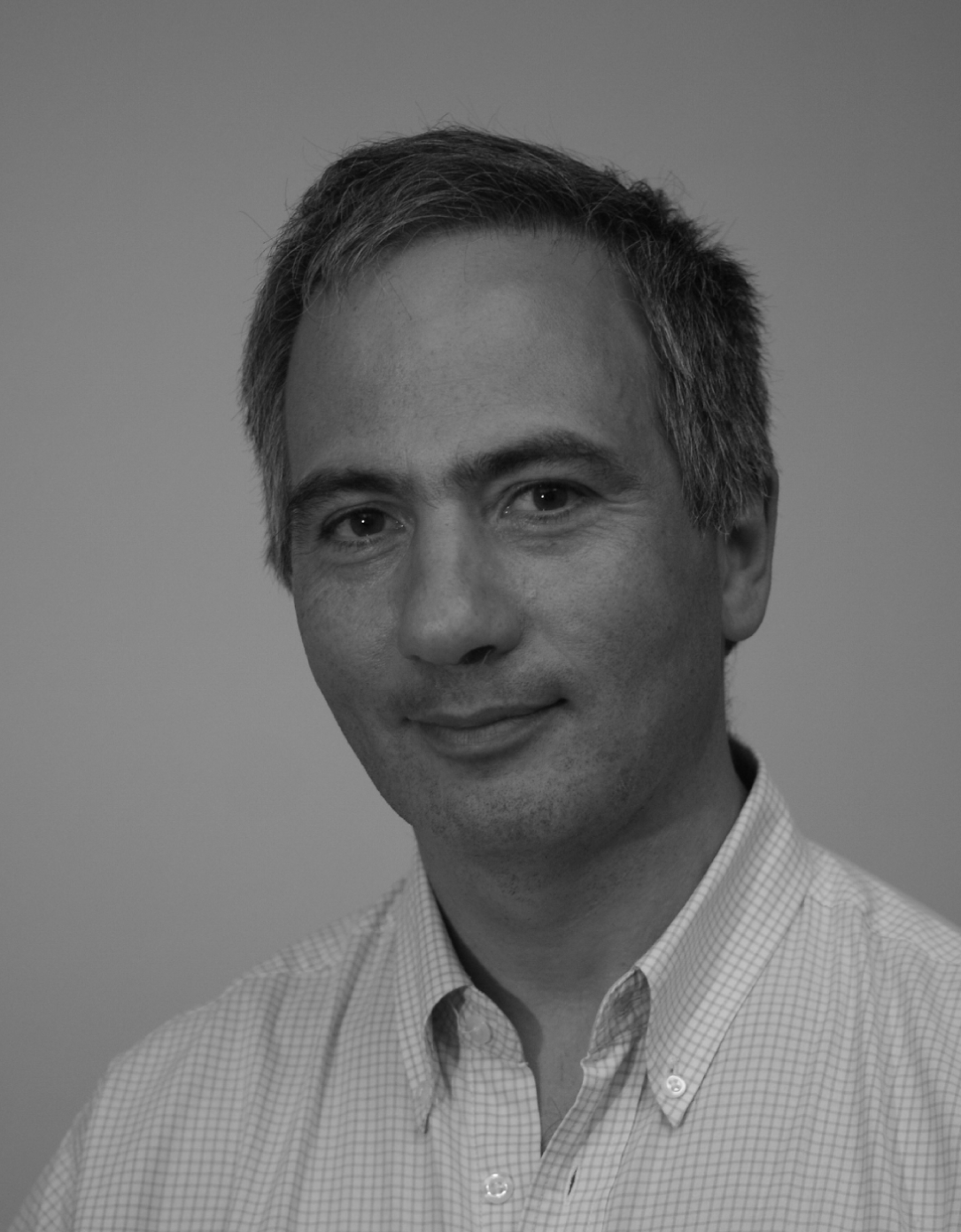}}]{Riccardo Leonardi} received the Diploma and the Ph.D.~degrees in Electrical engineering from the Swiss Federal Institute of Technology, Lausanne, Switzerland, in 1984 and 1987, respectively. He has been a researcher at UC Santa Barbara and Bell Laboratories from 1987 till 1991. Since 1992 he was appointed at the University of Brescia, Italy, where he established the Signal, Imaging, Networking and Communications (SINC) Group. He conducts research in signal and image representation for visual communications and visual content protection. He also is an expert in machine learning tools with application to multimedia content analysis, and medical imaging. Prof.~Leonardi is as an expert evaluator for the European Commission, and is a former President of the Italian Telecommunication and Information Technology Association (GTTI).
\end{IEEEbiography}

\begin{IEEEbiography}
[{\includegraphics[width=1in,height=1.25in,clip,keepaspectratio]{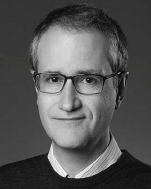}}]{Antonio Ortega} is Dean's Professor of Electrical and Computer Engineering at the University of Southern California (USC).  He received his undergraduate and doctoral degrees from Universidad Politecnica de Madrid, Madrid, Spain, and Columbia University, New York, NY. He is a fellow of the IEEE and EURASIP and currently serves as the VP of publications for the IEEE Signal Processing Society. He has received several paper awards, including the 2016 Signal Processing Magazine award. His recent research focuses on graph signal processing, machine learning, and multimedia compression. He is the author of the book, "Introduction to Graph Signal Processing," published by Cambridge University Press in 2022.  
\end{IEEEbiography}
\vfill

\end{document}